\documentclass[a4paper,final]{article}
\usepackage{hyperref}		% PDF hyperreferences??
\usepackage{listings}

\usepackage[utf8]{inputenc}
\usepackage[english,activeacute]{babel}
\usepackage{tabularx,wrapfig}
\usepackage{graphics}
\usepackage{graphicx}
\usepackage{epsfig}
\usepackage{amssymb,amsmath,amsthm,amsfonts}
\usepackage[all]{xy}
\usepackage{url}
\usepackage{subfigure}
\usepackage{chngpage}
\usepackage[figuresright]{rotating}
\usepackage{booktabs}
\usepackage{algorithm}
\usepackage{algorithmic}
\usepackage{comment}
\usepackage{morefloats}
\usepackage[shortlabels]{enumitem}
\newtheorem{theorem}{Theorem}

\newtheorem{definition}{Definition}
\newtheorem*{remark}{Remark}

\newcommand{\conecta}[3]{#1 \overset{#2}{\rightsquigarrow} #3}
\begin{document}

\title{Generalized Graph Pattern Matching}
\author{Pedro Almagro-Blanco and Fernando Sancho-Caparrini}
\maketitle

\section{Introduction}

Graphs are high expressive models and especially suitable for modelling highly complex structures, the number of applications that use them for modelling both their data and the processes that manipulate has grown dramatically in recent years.

The use of new conceptual structures in \textit{real-world} applications requires the development of new systems to store and query such structures in order to allow users to access data effectively and efficiently. As with all new technologies, it is still in the development phase and in search of a set of standards ensuring a continued growth.

The growing popularity of Graph Databases has led to the emergence of interesting problems related to storage and query in these systems. These databases have robust fundamentals in terms of basic information management (creation, access, deletion, and modification of individual elements of the structure), but they lack standards in some other tasks necessary for the storage and retrieval of information, like those related to more advanced query mechanisms.

Among the problems related to these query processes, the detection of patterns is considered as one of the fundamental ones since it includes other subproblems necessary to obtain powerful query systems, such as the search of subgraphs or minimum paths, or the study of connectivity \cite{distance-join, gupta2015neo4j}.

In this paper we present Generalized Graph Query (GGQ), a proposal to carry out queries in property graphs. GGQ represents a robust logical framework, making it especially useful in graph discovery procedures and generalizing other more basic tools that have been proved very useful on related tasks. In addition, we will present a collection of operations that allow to obtain complex GGQs from simpler ones, something very useful when automating the construction of complex queries.

\section{Graph Pattern Matching}

Graph Pattern Matching is an active research area since more than 30 years and its usefulness has been demonstrated in many areas, from artificial vision, to biology, through electronics, computer-aided design, and analysis of social networks, among others. Undoubtedly, its interest has grown even more as graph databases has become a tool for structuring and storing information in a transversal way in all areas of knowledge. For this reason, the problem of graph pattern matching expands, with slight variants, across different scientific communities, showing that is not a single problem defined under a common formalization, but a set of related problems.

The process by which we check the presence of a particular pattern in a particular set of data is called \textit{pattern detection}, and computationally represents the set of mechanical processes that allow to return one (or all) the \emph{occurrences} of the pattern.

The definition of what is meant by an \emph{occurrence} of a pattern in a graph varies according to how the pattern is defined. When it is given by a graph structure (although not necessarily using the same elementary sets of vertices and edges as the graph on which the query is made) it is usual to associate the occurrence with the existence of identifications (perhaps not as strong as isomorphisms) between the pattern and those subgraphs in the data that respect some imposed constraints \cite{gpmsocial}.

Some classifications that we can present in terms of the different possible ways to carry out the detection of patterns in graphs are: (a) \emph{Structural vs. Semantic}, (b) \emph{Exact vs. Inexact}, and (c) \emph{Optimal vs. Aproximate} \cite{matching}.

One other classification takes into account the relation to be met by the pattern regarding to the subgraphs that are considered its occurrences. It allows to divide the different techniques of graph pattern matching in those based on \emph{Isomorphisms} (an occurrence will be any subgraph that is isomorphic to the pattern), \emph{Graph Simulation} \cite{Milner} (an occurrence will be any subgraph for which there is a binary relation between the elements of the subgraph and the pattern which respects the types of the nodes and their adjacencies), \emph{Bounded Simulation} \cite{Fan, distance-join, Milner} (based on Graph Simulation but allowing to associate pattern edges to paths) and \emph{Regular Pattern Matching} \cite{rexp, phdthesis, Barcelo, Milner} (Bounded Simulation with regular expressions for pattern edges).

Current tools that allow querying patterns in graphs use either a declarative language (such as Cypher, SPARQL, or SQL), or an imperative language (with Gremlin as the clearest representative). In the case of declarative languages, it is responsibility of the system (ideally) to perform queries optimization. In the case of imperative languages, the execution plan is responsibility of the user, so they usually provide lower level approximations.

Next, we briefly present some query languages that have been used for graph pattern matching.

\emph{SQL} (Structured Query Language) is a declarative language for accessing Relational Database Management Systems (RDBMS). These types of databases were designed for tabular data with a fixed schema, and work best in contexts that are well defined from the beginning and where the records themselves are more important than the relationships between them. For this reason, trying to answer questions involving many relationships between data (as is usual in graph pattern queries) with a relational database involves numerous and costly operations between tables, that may become this option intractable. In spite of this, SQL has the expressive capacity necessary to be able to express most of graph pattern queries, and because relational databases have been the storage option chosen by most projects for decades, the first graph pattern query systems used SQL queries, such as \textit{Selection Graphs} that we will see later.

\emph{SPARQL} is a declarative query language for data stored in RDF format \cite{Segaran:2009:PSW:1696488} and is recognized as a key technology of the Semantic Web, its expressive ability to perform graph pattern matching is superior to that provided by SQL, but is still not intuitive for a human user. SPARQL allows structural, semantic, optimal, and exact queries based on subgraph isomorphism. Although SPARQL does not allow for any type of Graph Simulation or Regular Pattern Matching, language extensions such as PSPARQL \cite{alkhateeb2007rdf} have been developed to query RDF databases using patterns that make use of regular expressions. Thanks to the structure of the language, it is relatively easy to generate automatic queries in SPARQL and there are tools that use it as a final query language.

\emph{Gremlin} is, at the same time, a query language for databases and a graph oriented computation system. As a language, Gremlin is independent of the underlying database. Its syntax allows for declarative and imperative queries, and easily express complex queries on graphs. Each query consists of a sequence of steps that perform atomic operations on the data stream. Gremlin allows semantic, exact, optimal queries and is based on subgraph isomorphism. It is possible to design other query languages to compile to Gremlin language (eg, SPARQL can be compiled to run on Gremlin \footnote{\url{https://github.com/dkuppitz/sparql-gremlin}}).

\emph{Selection Graphs} are multi-relational patterns to express SQL based queries. They represent the foundation of the multi-relational decision tree induction algorithm MRDTL \cite{Leiva02mrdtl:a} as well as other multi-relational automatic learning algorithms \cite{DBLP:journals/corr/abs-1211-3871}. The purpose of selection graphs is to be used in top-down discovery procedures \cite{Knobbe99multi-relationaldecision}, thus operators to modify them through small changes are provided that allow the construction of complex queries in successive steps. As a counterpart, they have the disadvantage that these constructions can not contain cycles, limiting the expressiveness of queries. Selection graphs represent an exact, optimal type of graph pattern matching based on semantic graph isomorphism. In addition, being a graphical representation of SQL queries, they inherit the efficiency problems presented by the systems based on this technology. The new proposal we show in these pages can be considered as a generalization of some of the ideas that promoted this previous approach, avoiding some of their limitations. 

\emph{Cypher} is a declarative query language specifically developed to work on Neo4j graph database \footnote{\url{http://neo4j.org}}. Cypher is designed to be a \textit{human-friendly} query language, close to both developers and end users. Query patterns in graphs that are usually hard to express in other languages are very simple in Cypher \cite{cypher}, showing a high expressive capacity \cite{cypherpat}. Neo4j database closely follows the property graph model, but forces the edges to be typed. Unlike Gremlin, Cypher is not Turing complete, so it has some limitations, and it is higher level. Simple Cypher queries have good performance, but when queries follow complex patterns they can lead to a worst performance since it does not allow to indicate the application order of the different conditions. Cypher allows structural and semantic, optimal, exact, and based on subgraph isomorphism queries. In addition, it allows a type of Regular Pattern Matching in which the edges in the pattern are projected onto paths of the graph, and where constraints can be imposed through expressions that make use of the disjunction operator and the closure of Kleene. One limitation from the academic point of view is that it lacks an associated formal model, and some of its operations have not yet been validated. Despite this, because of its excellent expressiveness and acceptable performance, and because it consumes the data from a graph database with a widespread use, Cypher has been the election to implement the tests of GGQ.

Some other tools related to graph pattern matching are: \emph{GraphLog} \cite{graphlog}, that allows to structure the queries as graphs and evaluates the existence of patterns between a pair of nodes to generate a new edge between them; \emph{GraphQL} \footnote{\url{http://graphql.org/}}, a declarative query language developed by Facebook to allow external applications to access its information; \emph{Graql} \footnote{\url{https://grakn.ai}}, a declarative query language oriented to knowledge graphs; and \emph{GGQL} \cite{van2016pgql}, a SQL extension with additional graph pattern matching tools (analysis of accessibility between nodes, definition of paths, or construction of graphs).

\section{Preliminaries}

Given a set $V$, we denote:
$$V^0= \emptyset,\ V^1= V,\ V^{n+1} = V^n \times V,\ V^*=\bigcup_{n\geq 0}V^n$$

In general, we call \emph{sequences} or \emph{lists} the elements in $V^*$. If $x \in V^n$ then we say that $x$ has length $n$, and we write $|x|=n$.

If $x=(a_1,\dots,a_n),y = (b_1,\dots, b_m) \in V^*$, then the concatenation of $x$ and $y$ is the element of $V^*$ given by $ xy = (a_1, \dots, a_n, b_1, \dots, b_m) $.

For each $ x = (a_1, \dots, a_n) \in V^n $, we call \emph{support set of $x$} to $s(x) = \{a_i: 1 \leq i \leq n \} $. We write $ a \in x $ to indicate that $ a \in s(x) $.

For each $ a \in V $, we denote $ |a|_x = \#\{i: x_i = a \} $ (where $ \# (A) $ is the cardinal of the set $ A $), and we call \emph{multi-support of $ x $} to the set of pairs $ ms (x) = \{(a, | a |_x): a \in x \} $. We define the relation $\sim$, which can be easily proved to be equivalence in $V^n$, as: $x\sim y $ if and only if $ ms (x) = ms (y)$
And denote by $ V_\sim^n = V^n / \sim $ (set quotient of $ V^n $ under $ \sim $).

Our interpretation of these sets will be that $ V^n $ denotes the set of ordered tuples of $ V $, $ V_\sim^n $ denotes the set of unordered tuples of the same set (ie tuples in which elements are important, considering the possible repetitions, but not the order in which they appear).

Next, we present the \emph{Generalized Graph}, that covers the different variants of graph that can be found in the literature and that we will need when presenting our proposal of property graph pattern matching tool.

\begin{definition}
A \emph{Generalized Graph} is a tuple $ G = (V, E, \mu) $ where:
\begin{itemize}
\item $ V $ and $ E $ are sets, called, respectively, \emph{set of nodes} and \emph{set of edges} of $ G $.
\item $ \mu $ is a relation (usually we will consider it functional, but not necessarily) that associates each node or edge in the graph with its set of properties, that is, $ \mu: (V \cup E) \times R \rightarrow S $, where $ R $ represents the set of possible \emph{keys} for these properties, and $ S $ the set of possible \emph{values} associated.
\end{itemize} 

Usually, for each $\alpha \in R$ and $x\in V\cup E$, we write $ \alpha (x) = \mu (x,\alpha) $.

In addition, we require the existence of a special key for the edges of the graph, which we call \emph{incidences} and denote by $ \gamma $, which associates to each edge of the graph a tuple, ordered or not, of vertices of the graph.
\end{definition}

Although the definition that we have presented here is more general than those that can be found in the related literature, we will also call them \emph{Property Graphs}, since they suppose a natural extension of this type of graphs.

It should be noted that in generalized graphs, unlike traditional definitions, the elements in $ E $ are symbols representing the edges, and not pairs of elements from $V$, and $ \gamma $ is the function that associates to each edge the set of vertices that it connects.

\begin{definition}[Notation and definitions]
	In the context of the above definitions, we use the following notation:
\begin{itemize}
\item We usually identify $ e $ with $ \gamma(e) $. Thus, we interpret the edge as the collection of nodes that connects, as classic definitions of graphs.
\item Symmetrically, for every $ u \in V $ we write $ \gamma(u) = \{e  \in E: u  \in e \} $.
In general, a generalized graph may have a combination of directed and non-directed edges. If $ \gamma: E \rightarrow V^* $ we say $G$ to be \emph{Directed} (\emph{Not Directed}, otherwise).
\item For each $ e  \in E $, we define the \emph{arity} of $ e $ as $ \sum_{a  \in e} | a | _e $.
\item If $ \gamma: E \rightarrow V^2 \cup V_\sim^2 $ we say that the graph is \emph{Binary} (and it matches the most usual graph structure). Otherwise, the graph is a \emph{Hypergraph}.
\item An edge, $ e \in E $, is said to be a \emph{loop} if it connects a node to itself, that is, if it has arity other than 1 but $ s (e) $ is unitary.
\item An edge, $ e \in E $, is said to be \emph{incident} on a node, $ v  \in V $, if $ v  \in e $.
\item Two distinct nodes, $u,v\in V$ are called \emph{adjacent}, or \emph{neighbors}, in $ G = (V,E, \mu)$ if there exists $e\in E$ such that $\{u,v\}\subseteq e$.
\item If there are different edges in $ E $ with the same incidence, that is, edges connecting the same nodes, we will say that the graph is a \emph{Multi-graph}.
\item If $ e $ is a directed binary edge connecting $ u $ to $ v $, $ e = (u, v) $, we write $ u \overset{e}{\rightarrow} v $, and we also denote $e^o=u$ (\emph{output} of $e$) y $e^i=v$ (\emph{input} of $e$). In this case, for each $ u  \in V $ we write: $$\gamma^o(u)=\{e\in \gamma(u):\ e^o= u\}$$ $$\gamma^i(u)=\{e\in \gamma(u):\ e^i=u\}$$
which respectively denote the sets of \emph{outgoing edges} and \emph{incoming edges} of $ u $.
\item Given $ u  \in V$, we define the \emph{environment} of $ u $ in $ G $ as the set of nodes, including $ u$, that are connected to it, i.e.: $ \mathcal{N}(u) = \bigcup_{e  \in \gamma (u)} \gamma (e) $. When necessary, we will use the \emph{reduced environment} of $ u $, $ \mathcal{N}^* (u) = \mathcal{N} (u) \setminus u $.
\end{itemize}
\end{definition}

The notion of subgraph is obtained from the usual definition by imposing that the properties are also maintained in the common elements.

\begin{definition}
A subgraph of a graph $G = (V, E,\mu)$ is a graph $S = (V_S, E_S,\mu_S)$ such that $V_S \subseteq V$ and $E_S \subseteq E$ and $\mu_S\subseteq \mu_{|V_S \cup E_S}$. We denote $S \subseteq G$. 
\end{definition}

A fundamental concept when working with graphs is the concept of \emph{path}, which allows the study of distance relationships and connectivity conditions between different elements, extending the connectivity allowed by edges to more general situations.

As generalized graphs are considerably more general than the usual ones we should give some notions about the position of a node in an edge:

\begin{definition}
If $e\in E$ and $\gamma(e)=(v_1,\dots,v_n)\in V^n$, then for each $v_i\in s(e)$ we define its \emph{order} in $e$ as $ord_e(v_i)=i$. If $e\in V_\sim^n$, then for each $v\in s(e)$ we define $ord_e(v)=0$.

We denote $u\leq_e v$ to indicate that $ord_e(u)\leq ord_e(v)$.
\end{definition}

From this relation of order between the nodes that an edge connects, we can define what we mean by a path in a graph.

\begin{definition}
Given a graph $G=(V,E,\mu)$, the set of \emph{paths} in $G$, denoted by $\mathcal{P}_G$, is defined as the minimal set verifying:
\begin{enumerate}
	\item If $e\in E$, $u,\ v\in e$ with $u \leq_e v$, then $\rho=u\overset{e}{\rightarrow}v\in \mathcal{P}_G$, and $sop_V(\rho)=(u,v)$, $sop_E(\rho)=(e)$. We will say that $\rho$ \emph{connects} the nodes $u$ and $v$ of $G$, and we will denote it by $\conecta{u}{\rho}{v}$.
	\item If $\rho_1,\ \rho_2\in \mathcal{P}_G$, with $\conecta{u}{\rho_1}{v},\ \conecta{v}{\rho_2}{w}$, then $\rho_1\cdot\rho_2 \in \mathcal{P}_G$, with $\conecta{u}{\rho_1\cdot\rho_2}{w}$, $sop_V(\rho_1\cdot\rho_2)=sop_V(\rho_1)sop_V(\rho_2)$, $sop_E(\rho_1 \cdot \rho_2)=sop_E(\rho_1)sop_E(\rho_2)$.
\end{enumerate}
When $ u = v $ we say that $ \rho $ is a \emph{closed} path, and if there are not repeated edges in $ \rho $ we say that it is a \emph{cycle}.
\end{definition}

If $\rho\in\mathcal{P}_G$, with $sop_V(\rho)=(u_1,\dots,u_{n+1})$ and $sop_E(\rho)=(e_1\dots,e_n)$, then we write:
$$\rho=u_1 \overset{e_1}{\rightarrow} u_2 \overset{e_2}{\rightarrow} \dots \overset{e_n}{\rightarrow} u_{n+1}$$

Generally we will write $u\in \rho$ to express that $u\in sop_V(\rho)$, and $e\in \rho$ to express that $e\in sop_E(\rho)$. 

\begin{remark}
	\quad 
	\begin{itemize}
	\item Following a similar notation to the case of directed binary edges, if $\rho\in \mathcal{P}(G)$ and $\conecta{u}{\rho}{v}$, then we write $\rho^o=u$ and $\rho^i=v$.
	\item When necessary, we denote the paths through $u$, starting in $u$, and ending in $ u$, respectively, by:
	$$\mathcal{P}_u(G)=\{\rho \in \mathcal{P}(G):\ u\in \rho\}$$
	$$\mathcal{P}_u^o(G)=\{\rho \in \mathcal{P}(G):\ \rho^o=u\}$$
	$$\mathcal{P}_u^i(G)=\{\rho \in \mathcal{P}(G):\ \rho^i=u\}$$
\end{itemize}
\end{remark}

\section{Generalized Graph Query}

Next, we present \emph{Generalized Graph Query} (GGQ, for short), our proposal to perform graph pattern matching on generalized graphs. Taking into account the different classifications mentioned above, we can say that this proposal allows to carry out structural and semantic, exact, optimal, and based on a type of Regular Pattern Matching queries, allowing the edges of the pattern to be projected on paths (not necessarily edges). Also, GGQ allows to express more complex constraints on each element of the pattern and perform cyclic queries.

One of the characteristics that we pursue for our tool is to provide a mechanism to obtain complementary patterns to a given one. This means that if a structure does not verify a pattern it must always verify one of its complementary patterns. As we have seen in previous section, many of the tools developed to perform queries of patterns in graphs require for a projection to be fulfilled between the pattern and the structure to be evaluated. This projection prevents us from evaluating the non existence of elements, something that we will need to generate these complementary patterns, for this reason our proposed matching system will not make use of projections, but is based on logical predicates, facilitating the generation of complementary patterns.

We want to emphasize that one of our main goals is to provide a complete formalization of the model, but with the secondary objective of providing an implementation that is usable from a practical point of view \footnote{\url{https://github.com/palmagro/ggq}} (as a proof of concept more than as a professional tool in this first stage).

In pursuit of our objectives, we will rely on Selection Graph model, extending it to add Regular Pattern Matching and some additional features that will allow us to obtain a greater expressiveness power in the patterns.

As main differences with the query systems from the previous section we can indicate that:

\begin{itemize}
    \item GGQ may contain cycles. It will be a later problem to consider implementations of GGQs that handle cycles properly, considering additional constraints to ensure certain levels of efficiency in their actual execution, or being careful when designing the query to create a pattern that is efficient in the available implementation.

    \item GGQ can evaluate subgraphs. In selection graphs it is only possible to evaluate a single node representing the target table. In the case of GGQ, fixed elements (elements that must belong to the subgraph under evaluation) will be represented through a predicate that forces them to be contained in the subgraph to be evaluated.

    \item Individual edges of the GGQ can be projected onto paths in the graph where the pattern is being checked. For this reason, predicates will be used in a similar way as in Regular Pattern Matching.

    \item The predicates associated with nodes or edges in the GGQ can evaluate structural and semantic characteristics beyond the properties stored through the $ \mu $ function (for example, using metrics on the graph or its elements).

\end{itemize}

We will briefly formalize what we understand concretely by a predicate defined on a graph.

Consider $ \Theta $, a collection of function, predicate, and constant symbols, containing all the functions from $ \mu $ together with constants associated with each element of the graph and possibly some additional symbols (for example, metrics defined on the elements of the graph). From this set of symbols we can define a First Order Language with equality, $ L $, making use of $ \Theta $ as a set of non logical symbols, on which we construct, in the usual way, the set of terms of the language and the set of formulas, $ FORM (L) $, which we will call \emph{predicates}.

Although, generally, the definable formulas in $ L $ can be applied to all objects in the universe, which in our context will consist in elements of graphs (nodes, edges, and structures formed from them), when we want to make explicit the types of objects we are working with, we can write $ FORM_V (L) $ for formulas on nodes, $ FORM_E (L) $ for formulas on edges, $ FORM_\mathcal{P}(L) $ for formulas on paths, etc.

In order to simplify the notation, when there is no possibility of confusion we will use $ FORM $ to denote $ FORM (L) $.

In addition, and taking advantage of the expressiveness capacity of generalized graphs, we define the query system using the same structure:

\begin{definition}
    A \emph{Generalized Graph Query (GGQ)} over $L$ is a binary generalized graph, $Q = (V_Q,E_Q,\mu_Q)$, where exist $\alpha$ and $\theta$, properties in $\mu_Q$, such that:
    \begin{itemize}
        \item $\alpha:V_Q\cup E_Q\rightarrow \{+,-\}$ total.
        \item $\theta:V_Q\cup E_Q\rightarrow FORM(L)$ associates a binary predicate, $\theta_x$, to each element $x$ of $V_Q\cup E_Q$.
    \end{itemize}
\end{definition}

We will write $ Q \in GGQ(L) $ to denote that $ Q $ is a Generalized Graph Query over $ L $ (if the language is prefixed, we simply write $ Q \in GGQ $).

In the semantics associated with a GGQ we will use the second input of these binary predicates to impose requirements of membership on subgraphs of $ G $ (the general graph on which we are evaluating queries), whereas the first input must receive elements of the corresponding type to which it is associated: if $S$ is a subgraph and $a\in V_Q$ then $\theta_a(.,S)\in FORM_V$, and if $e\in E_Q$ then $\theta_e(.,S)\in FORM_\mathcal{P}$. For example:
\begin{align*}
\theta_a(v,S)&=\exists z\in S\ (\conecta{z}{}{v})\\
\theta_e(\rho,S)&=\exists y,z (\conecta{y}{\rho}{z} \wedge y \notin S\wedge z\in S)
\end{align*}

$\theta_a(v,S)$ will work for nodes, and it will be verified when there is a path in $ G $ that connects a node of $ S $ (the subgraph we are evaluating) with $ v $, the input node on which it is evaluated. $\theta_e(\rho,S)$ will work for paths, and will be verified when the evaluated path, $ \rho $, connects $ S $ with its complementary (in $ G $).

Given a GGQ under the above conditions, $ x^+ $, respectively $ x^- $, will indicate that $ \alpha (x) = + $, respectively $ \alpha (x) = - $, and $V_Q^+/V_Q^-$ (respectively, $E_Q^+/E_Q^-$) the set of positive/negative nodes (respectively, edges). If for an element, $ \theta_x $ is not explicitly defined, we assume it to be a tautology (generally denoted by $T$).

As we will see below, positive elements of the pattern represent elements verifying the associated predicates that must be present in the graph, while negative ones represent elements that should not be present in the graph.

In order to be able to express more easily the necessary conditions that define the application of a GGQ on a graph, as well as the results that we will see later, we introduce the following notations:

\begin{definition}
    Given a GGQ, $Q=(V_Q,E_Q,\mu_Q)$, the set of $Q$-predicates associated to $Q$ is:
    \begin{enumerate}
        \item For each edge, $e\in E_Q$, we define:
        $$Q_{e^o}(v,S)=\exists \rho\in \mathcal{P}_v^o(G)\ \left(\theta_e(\rho,S)\wedge \theta_{e^o}(\rho^o,S) \wedge \theta_{e^i}(\rho^i,S)\right)$$
        $$Q_{e^i}(v,S)=\exists \rho\in \mathcal{P}_v^i(G)\ \left(\theta_e(\rho,S)\wedge \theta_{e^o}(\rho^o,S) \wedge \theta_{e^i}(\rho^i,S)\right)$$

        In general, we will write $Q_{e^*}(v,S)$, where $*\in \{o,i\}$, and we will denote:
        $$Q_{e^*}^+ = Q_{e^*},\hspace{1cm} Q_{e^*}^- = \neg Q_{e^*}$$
        \item For each node, $n\in V_Q$, we define:
        \begin{align*}
        Q_n(S)&=\exists v\in V\ \left(\bigwedge_{e\in \gamma^o(n)} Q_{e^o}^{\alpha(e)}(v,S)\ \wedge \bigwedge_{e\in \gamma^i(n)} Q_{e^i}^{\alpha(e)}(v,S)\right)\\
        & = \exists v\in V\ \left(\bigwedge_{e\in \gamma^*(n)} Q_{e^*}^{\alpha(e)}(v,S)\right)
        \end{align*}
        That can be written generally as:
        $$Q_n(S)=\exists v\in V\ \left(\bigwedge_{e\in \gamma(n)} Q_{e}^{\alpha(e)}(v,S)\right)$$
        In addition, we denote:
        $$Q_n^+ = Q_n,\hspace{1cm} Q_n^- = \neg Q_n$$
    \end{enumerate}
\end{definition}

From these notations, we can formally define when a subgraph matches a given GGQ:

\begin{definition}
\label{verifica}
    Given a subgraph $S$ of a property graph, $G=(V,E,\mu)$, and a Generalized Graph Query, $Q=(V_Q,E_Q,\mu_Q)$, both over language $L$, we say that $S$ \emph{matches} $Q$, and we will denote $S\vDash Q$, if the next formula is verified:

    $$Q(S)=\bigwedge_{n\in V_Q} Q_n^{\alpha(n)}(S)$$

    Otherwise, we will write: $S\nvDash Q$.
\end{definition}

\begin{figure}[h!]
    \begin{center}
        \includegraphics[scale=0.4]{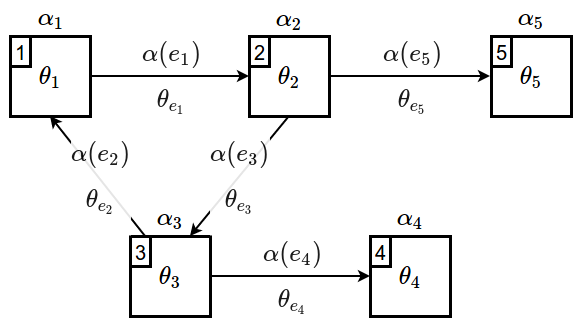}
    \end{center}
    \caption{%
        Generalized Graph Query Example.
    }%
    \label{abstractpqg}
\end{figure}

Note that, in particular, using $ S = G $ we can define when a graph matches a GGQ. A generic GGQ example is shown in Figure \ref{abstractpqg}.

One of the objectives for GGQ is to provide power enough to express conditions that make use of elements that are outside the subgraph being evaluated, something that has been proven necessary to have a powerful query language and, except in the selection graphs (and only to a very limited extent), it is not present in the other previous solutions.

Although the definition of GGQ makes use of binary graphs (not hypergraphs), since it projects edges on paths connecting pairs of nodes, the generalized graph concept is flexible enough to allow other interpretations where GGQ can use more general structures. In addition, and it is important to emphasize this fact, although GGQ are binary graphs, they can be applied on hypergraphs with properties, since the concept of path that connects pairs of nodes is defined independently of the arity of the intervening edges. In these cases, a somewhat more complex notation should be used to define the $Q$-predicates, but it is completely feasible. For the sake of simplicity, and because of the lack of hypergraph databases, we have restricted the definitions to these particular cases, but they are open to be extended to more general cases at the moment in which the use of hypergraphs is generalized as tools of modelling and storage, since in most of the (rare) occasions they have been needed, the problem has always been solved by means of the creation of new types of nodes and binary edges that simulate the presence of hyperlinks.

Before going on to analyse some interesting properties about GGQ and how to construct them, let's see some examples that allow us to understand how they are interpreted and the expressive capacity they allow.

\section{Some Representative Examples}

\begin{figure}[h!]
    \begin{center}
        \includegraphics[width=\textwidth]{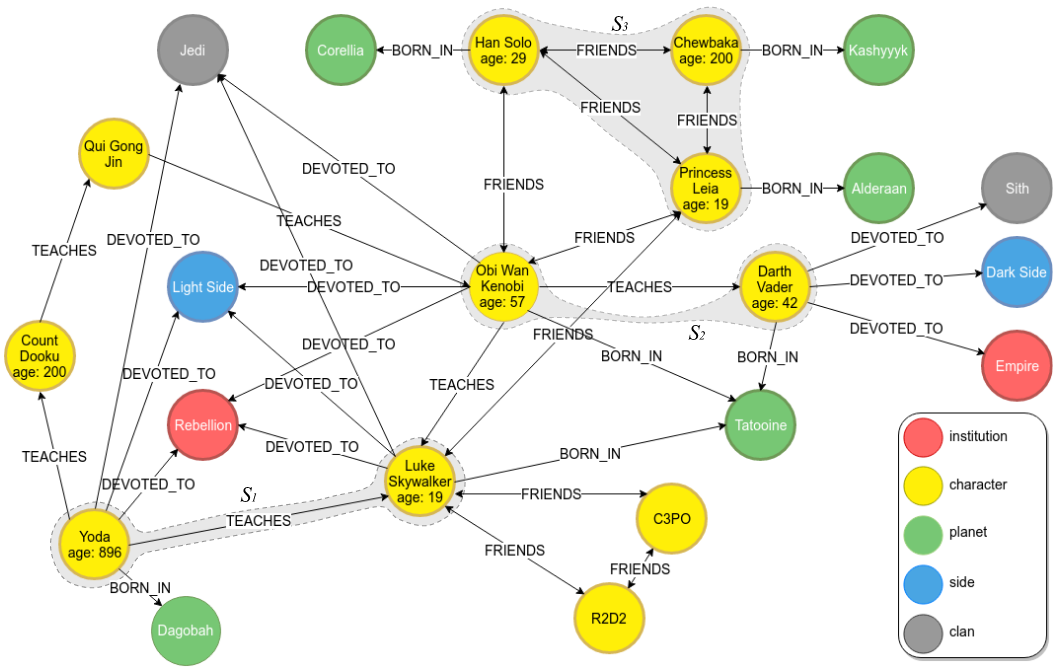}
    \end{center}
    \caption{%
        Starwars graph to present some Generalized Graph Query examples.
    }%
    \label{starwars}
\end{figure}

Figure \ref{starwars} shows a property graph that corresponds to a section of a graph database that contains information about the main characters of the Starwars series and that is frequently used as a simple example to do demonstrations related to graph database capabilities \footnote{http://console.neo4j.org/?id=StarWars}. We will use this graph to present some GGQ examples and to check the verification of some specific subgraphs.

In order to simplify the representation of queries and subgraphs, one of the properties in $ \mu $, which we will call $ \tau $ and which represents types on nodes and edges, will be expressed directly on the edges and, in the case of nodes, through colours. In addition, the \textit{name} property of the nodes will be represented directly on them, and the undirected edges will be represented as bidirectional edges.

\begin{figure}[h!]
    \begin{center}
        \includegraphics[scale=0.35]{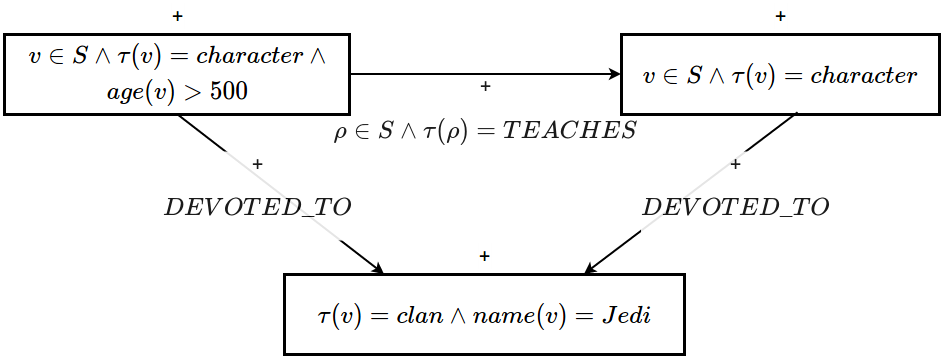}
    \end{center}
    \caption{%
        Example 1 GGQ.
    }%
    \label{pqg1}
\end{figure}

\begin{figure}[h!]
    \begin{center}
        \includegraphics[scale=0.35]{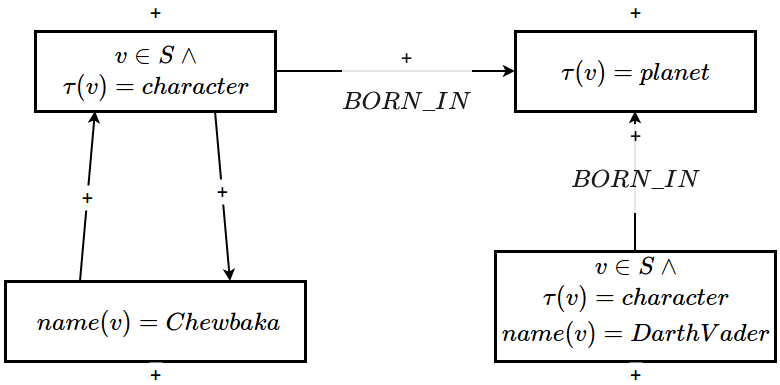}
    \end{center}
    \caption{%
        Example 2 GGQ.
    }%
    \label{pqg2}
\end{figure}

\begin{figure}[h]
    \begin{center}
        \includegraphics[scale=0.35]{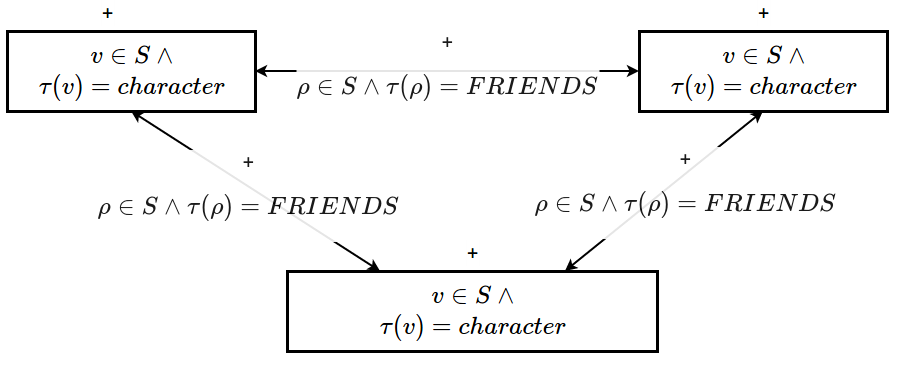}
    \end{center}
    \caption{%
       Example 3 GGQ.
    }%
    \label{pqg3}
\end{figure}

\begin{figure}[h]
    \begin{center}
        \includegraphics[scale=0.35]{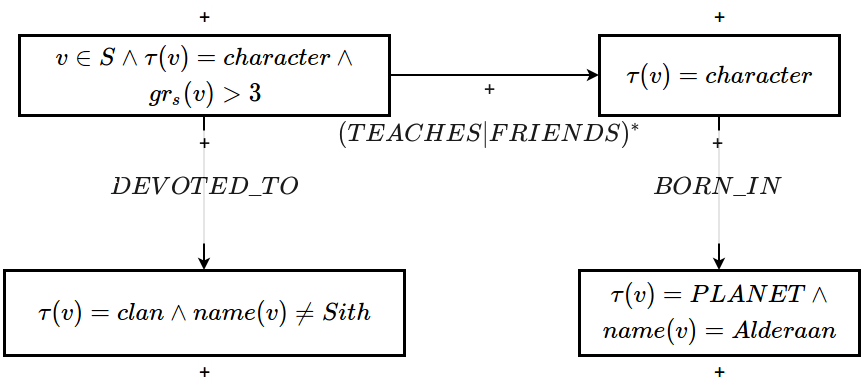}
    \end{center}
    \caption{%
        Example 4 GGQ.
    }%
    \label{pqg4}
\end{figure}

\begin{figure}[h]
    \begin{center}
        \includegraphics[scale=0.35]{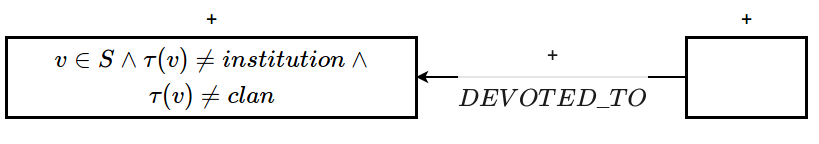}
    \end{center}
    \caption{%
       Example 5 GGQ.
    }%
    \label{pqg5}
\end{figure}

\begin{figure}[h]
    \begin{center}
        \includegraphics[scale=0.35]{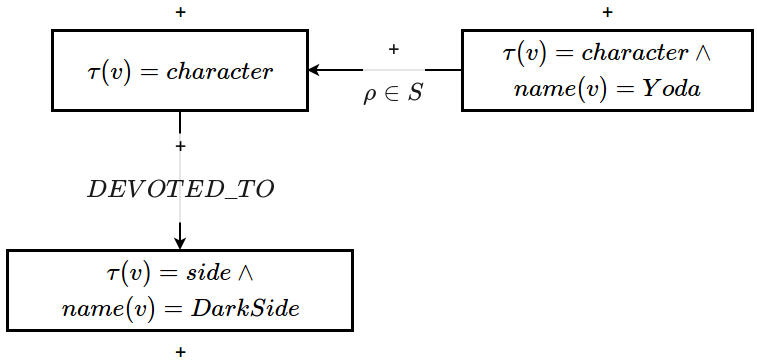}
    \end{center}
    \caption{%
        Example 6 GGQ.
    }%
    \label{pqg6}
\end{figure}

The graphical representation of the example GGQs are shown in figures \ref{pqg1} to \ref{pqg6}. When analysing the interpretation of these queries we will also indicate some subgraphs of $ G $ verifying them. The $ \alpha $ property will be directly represented on the elements of GGQ by means of a $ + / - $ symbol, and $ \theta $ property directly in the element (but in the case of tautologies). In expressions of the type $ \tau (\rho) = X $ in the predicate of an edge, $ X $ is interpreted as a regular expression that must be verified by the sequence of $ \tau $ properties of $ sop_E (\rho) $.

GGQ $ P_1 $ (Figure \ref{pqg1}) can be interpreted in natural language through the following sentence: \textit{Characters and student-teacher relationship in which both are devout of the Jedi and the teacher has more than 500 years}. In this case, structural constraints are imposed through the presence of edges and through predicates that make use of $ \tau $, \textit{name}, and \textit{age} properties. This GGQ will be verified by subgraphs where two nodes and one edge connecting them can be projected (the three elements marked as positive elements in the GGQ) verifying the imposed requirements. If there is a character who has taught himself (which would be given by a TEACHES loop) that is more than 500 years old and devout of the Jedi, any subgraph containing this node would also match this pattern. Subgraph marked as $ S_1 $ in Figure \ref{starwars} matches $ P_1 $.

GGQ $ P_2 $ (Figure \ref{pqg2}) can be interpreted in natural language through the following sentence: \textit{Subgraphs containing Darth Vader and a character coming from the same planet and connected to Chewbaka}. This GGQ presents a positive node representing the character \textit{Chewbaka} (using $name$) and imposes a constraint requiring that one of the nodes in the evaluated subgraph has to be connected to it. Subgraph marked as $ S_2 $ in Figure \ref{starwars} matches $ P_2 $.

GGQ $ P_3 $ (Figure \ref{pqg3}) presents a cycle using \texttt{FRIENDS} relations, and $ S_3 $ (highlighted in Figure \ref{starwars}) is a subgraph that verifies it. Any subgraph containing three characters that are friends with each other will verify $ P_3 $, for example, subgraphs containing the cycle formed by \texttt{Luke Skywalker}, \texttt{R2D2} and \texttt{C3PO}, or by \texttt{Han Solo}, \texttt{Princess Leia}, and \texttt{Chewbaka}.

GGQ $P_4$ (Figure \ref{pqg4}) can be interpreted in natural language through the following sentence: \textit {Character with an outgoing degree greater than 3, devout of a clan other than Sith and that is connected through \texttt{FRIENDS} or \texttt{TEACHES} relationships with someone who comes from Alderaan}. In this case, a regular expression has been used to express a path composed of FRIENDS or TEACHES type relationships, and an auxiliary function, $ gr_s (v) $, has been used to refer to the outgoing degree of the $ v $ node. Any subgraph containing \texttt{Luke Skywalker} node or \texttt{Obi Wan Kenobi} node will verify $ P_4 $.

GGQ $ P_5 $ (Figure \ref{pqg5}) represents the query: \textit{Nodes that are not institutions or clans, but have devotees}, and is only verified by nodes of type \textit{side}. In this case, a node whose property $ \theta $ is a tautology (represented as an empty node) has been used.

GGQ $ P_6 $ (Figure \ref{pqg6}) could be interpreted as: \textit{Paths connecting Yoda with characters from the Dark Side}. Any subgraph containing the path (\texttt{Yoda}) $ \rightarrow $ (\texttt{Count Dooku}) $ \rightarrow $ (\texttt{Qui Gong Jin}) $ \rightarrow $ (\texttt{Obi Wan Kenobi}) $ \rightarrow $ (\texttt{Darth Vader}) will verify $ P_6 $.

\section{Refinement Sets}

In previous sections we have seen that GGQ can be interpreted as predicates over the family of subgraphs of a prefixed graph $G$. It would be interesting to obtain computationally effective ways for building complex GGQ by using basic operations to obtain families of predicates to automatically analyse the structure of subgraphs of $G$. In this section we will give a first approximation to a constructive method that is useful to perform this type of tasks.

\begin{definition}
    Given $Q_1,\ Q_2\in GGQ$, we will say that $Q_1$ \emph{refines} $Q_2$ in $G$, denoted as $Q_1\preceq_G Q_2$ (simply $\preceq$ when working with a prefixed graph) if:

    $$\forall S\subseteq G\ (S\vDash Q_1 \Rightarrow S\vDash Q_2)$$
\end{definition}

\begin{figure}[h]
    \begin{center}
        \includegraphics[scale=0.33]{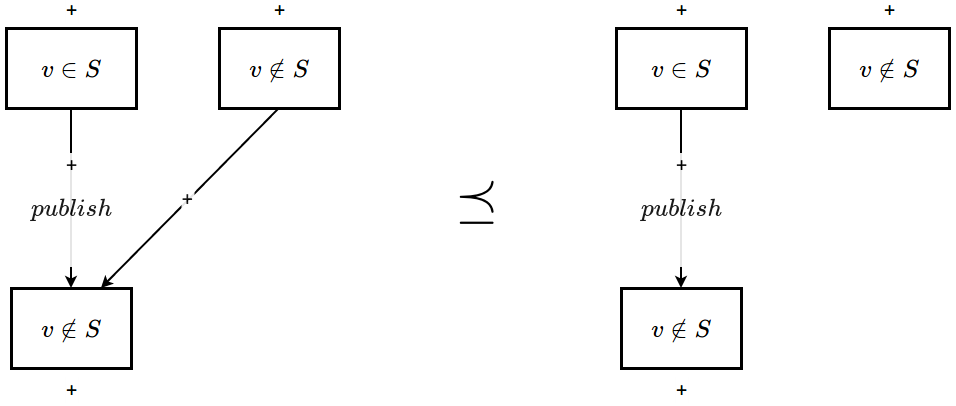}
    \end{center}
    \caption{%
        GGQ Refinement.
    }%
    \label{ejemplo-refinar}
\end{figure}

An example is shown in Figure \ref{ejemplo-refinar}. The GGQ on the left refines the GGQ on the right because in addition to requiring one of the nodes in the subgraph under evaluation to be connected to a node that does not belong to that subgraph through a \texttt{publish} relationship, it is also required that the target node of that relationship has an incoming edge that starts from a node that does not belong to the subgraph under evaluation.

In a natural way, we can define when two GGQ are equivalent as queries, which will happen when both are verified exactly by the same subgraphs.

\begin{definition}
    Given $Q_1,\ Q_2\in GGQ$, we will say that they are \emph{equivalents} in $G$, denoted by $Q_1\equiv_G Q_2$ (simply $\equiv$ when working on a prefixed graph) if:

    $$Q_1\preceq_G Q_2\ \wedge \ Q_2\preceq_G Q_1$$
\end{definition}

\begin{figure}[h]
    \begin{center}
        \includegraphics[scale=0.33]{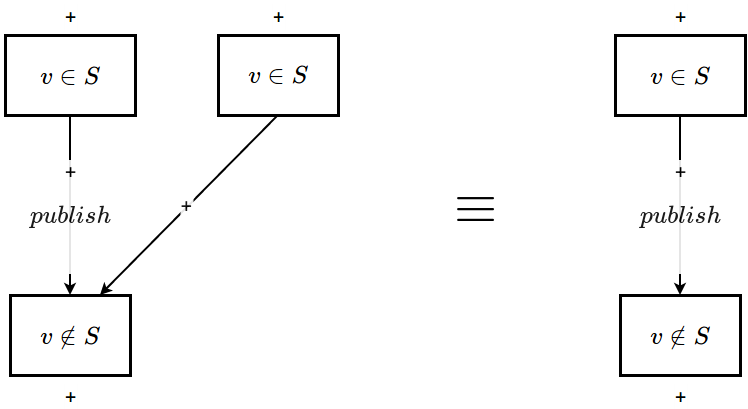}
    \end{center}
    \caption{%
        GGQ equivalence.
    }%
    \label{ejemplo-equivalencia}
\end{figure}

Figure \ref{ejemplo-equivalencia} shows an example of equivalence. Note that the only structural difference between the two GGQ is the existence, in the left one, of a node belonging to the subgraph to be evaluated that has an output edge whose destination is a node that is not included in the subgraph under evaluation. This restriction is included in the GGQ on the right, since there is a node in the subgraph under evaluation with an edge of type \texttt{publish} connected to a node not belonging to that subgraph.

It is easy to prove the following result, which tells us that $ \preceq_G $ generates a partial order relation on the GGQ considering equivalence as equality (working in the quotient space that determines equivalence).

\begin{theorem}
    For every property graph, $G$, $(GGQ, \preceq_G)$ is a partial ordered set. That is:
    \begin{enumerate}
        \item $\forall Q\in GGQ\ (Q\preceq_G Q)$.
        \item $Q_1\preceq_G Q_2\ \wedge \ Q_2\preceq_G Q_1 \Rightarrow Q1\equiv_G Q2$.
        \item $Q_1\preceq_G Q_2\ \wedge \ Q_2\preceq_G Q_3 \Rightarrow Q_1\preceq_G Q_3$.
    \end{enumerate}
\end{theorem}

It is easy to verify that, in general, $ \preceq_G $ does not generate a total order relation. To do this, just find two GGQ, $ Q_1 $ and $ Q_2 $, for which $ (Q_1 \rightarrow Q_2) \lor (Q_2 \rightarrow Q_1) $ is not met. For example, if $ Q_1 $ is a GGQ composed of a single positive node with the constraint $ v \in S $, and $ Q_2 $ is a GGQ also composed of a single positive node with the constraint $ v \notin S $, then $ Q_1 $ will require that the subgraph under evaluation is not empty, and $ Q_2 $ will require that there be some node in the graph that does not belong to the subgraph under evaluation. Both constraints are independent so there is no implication between the predicates represented by these GGQ.

Next, we analyze the relationship between the topological structure of a GGQ and its functionality as a predicate on subgraphs. In general, it is hard trying to extract logical properties of the predicate from the structural properties of the graph representing it, but we can obtain some useful conditions that will allow us to constructively manipulate the structures to modify the interpretation of the GGQ in a controlled way.

\begin{definition}
    Given $Q_1,\ Q_2\in GGQ$, we say that $Q_1$ is a \emph{$Q^-$-conservative extension} of $Q_2$, and we will denote it by $Q_2\subseteq^- Q_1$, if:
    \begin{enumerate}
        \item $Q_2\subseteq Q_1$ (as generalized graphs, so in the $ Q_2 $ elements the values of $ \alpha $ and $ \theta $ should coincide for both GGQ).
        \item For each negative node in $Q_2$, $n\in V_{Q_2}^-$, and every edge incident to it in $Q_1$, $e \in \gamma_{Q_1}(n)$, exists an edge incident to it in $Q_2$, $e'\in \gamma_{Q_2}(n)$, imposing the same restriction, that is: $Q_e\equiv Q_{e'}$.
    \end{enumerate}
\end{definition}

\begin{figure}[h]
    \begin{center}
        \includegraphics[scale=0.33]{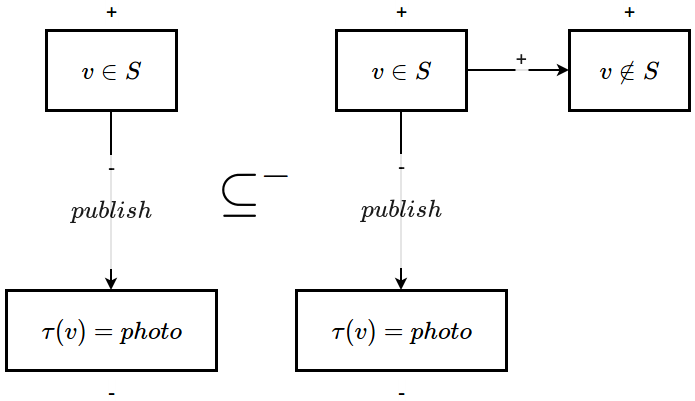}
    \end{center}
    \caption{%
        $Q^-$-conservative extension.
    }%
    \label{ejemplo-conservativa}
\end{figure}

Figure \ref{ejemplo-conservativa} shows an example of a conservative $Q^-$-extension. The extension made in the GGQ on the left to get the GGQ on the right imposes new restrictions on the positive node but does not add any further restrictions to the negative node.

Since negative nodes add non-existence constraints to subgraph verification, conservative $ Q^-$-extensions ensure that new constraints are not being added to them. Hence, we can give the next result:

\begin{theorem}
    Given $Q_1,Q_2\in GGQ$, if $Q_2\subseteq^- Q_1$ then $Q_1\preceq Q_2$.
\end{theorem}

\begin{proof}
Since $Q$-predicates associated to edges depend only on the information in the edge itself (which considers the value of $\theta$ at its incident nodes, regardless of the value of $ \alpha $ in them), we can state:
    $$\forall e\in E_{Q_2}\ ({Q_1}_e^{\alpha(e)}={Q_2}_e^{\alpha(e)})$$

Considering this fact, we analyse how the $Q$-predicates associated with the nodes for both GGQ behave:
    \begin{itemize}
        \item If  $n\in V_{Q_2}^-$, since $Q_2\subseteq^- Q_1$, is trivial that ${Q_1}_n^-={Q_2}_n^-$.
        \item If $n\in V_{Q_2}^+$, then ${Q_1}_n^+\rightarrow {Q_2}_n^+$, because (we will note $\gamma_1$, $\gamma_2$ the incidence functions of $Q_1$ and $Q_2$, respectively):
        \begin{align*}
        {Q_1}_n^+   &= \exists v\in V\ \left( \bigwedge_{e\in \gamma_1(n)} {Q_1}_e^{\alpha(e)}\right)\\
                &= \exists v\in V\ \left( \bigwedge_{e\in \gamma_1(n)\cap E_{Q_2}} {Q_1}_e^{\alpha(e)}\ \wedge \ \bigwedge_{e\in \gamma_1(n)\smallsetminus E_{Q_2}} {Q_1}_e^{\alpha(e)}\right)\\
                &= \exists v\in V\ \left( \bigwedge_{e\in \gamma_2(n)\cap E_{Q_2}} {Q_2}_e^{\alpha(e)}\ \wedge \ \bigwedge_{e\in \gamma_1(n)\smallsetminus E_{Q_2}} {Q_1}_e^{\alpha(e)}\right)\\
                &\rightarrow {Q_2}_n^+
        \end{align*}
    \end{itemize}
    Hence:
    \begin{align*}
    Q_1 &= \bigwedge_{n\in V_{Q_1}} {Q_1}_n^{\alpha(n)} = \bigwedge_{n\in V_{Q_2}} {Q_1}_n^{\alpha(n)} \ \wedge \ \bigwedge_{n\in V_{Q_1}\smallsetminus V_{Q_2}} {Q_1}_n^{\alpha(n)}\\
    &= \bigwedge_{n\in V_{Q_2}^+} {Q_1}_n^{\alpha(n)} \ \wedge \ \bigwedge_{n\in V_{Q_2}^-} {Q_1}_n^{\alpha(n)} \ \wedge \ \bigwedge_{n\in V_{Q_1}\smallsetminus V_{Q_2}} {Q_1}_n^{\alpha(n)}
\end{align*}
    \begin{align*}
    &\rightarrow \bigwedge_{n\in V_{Q_2}^+} {Q_2}_n^{\alpha(n)} \ \wedge \ \bigwedge_{n\in V_{Q_2}^-} {Q_2}_n^{\alpha(n)} \ \wedge \ \bigwedge_{n\in V_{Q_1}\smallsetminus V_{Q_2}} {Q_1}_n^{\alpha(n)}\\ 
    &= \bigwedge_{n\in V_{Q_2}} {Q_2}_n^{\alpha(n)} \ \wedge \ \bigwedge_{n\in V_{Q_1}\smallsetminus V_{Q_2}} {Q_1}_n^{\alpha(n)}\\
    &\rightarrow {Q_2}
    \end{align*}
\end{proof}

Previous result suggests that a GGQ can be refined by adding nodes (of any sign) and edges to the existing positive nodes, but because of the (negated) interpretation of $ Q $-predicates associated with negative nodes, care must be taken to maintain their environment to be sure that adding more edges does not weaken the imposed conditions (and, therefore, we would not get refined predicates).

In order to obtain controlled methods of query generation, in the following we will give a constructive method to refine GGQ by unit steps. To do this, we will start by looking at how GGQ behaves when cloning nodes.

A clone consists of making copies of existing nodes, and cloning all the edges incident on them (and between them, in case we clone several nodes that are connected in the original GGQ). Of course, the cloning operation can be done on any generalized graph, not only on GGQ.

\begin{definition}
    Given a generalized graph $G=(V,E,\mu)$, and $W\subseteq V$, we define the \emph{clone of $G$ by duplication of $W$}, denoted by $Cl_G^W$, as:

    $$Cl_G^W=(V\cup W',\ E\cup E',\mu\cup \{(n',\mu(n))\}_{n\in W}\cup \{(e',\mu(e))\}_{e'\in E'})$$
    where:
    \begin{itemize}
        \item for each $n\in W$, $n'$ is a new node, and $W'=\{n'\ :\ n\in W\}$, 
        \item $ E '$ is a set of new edges obtained from incident edges on nodes of $ W $ where nodes of $ W $ are replaced by copies of $ W' $ (edges connecting original nodes with cloned nodes, and edges connecting cloned nodes, are cloned).
    \end{itemize}
\end{definition}

\begin{figure}[h]
    \begin{center}
        \includegraphics[scale=0.33]{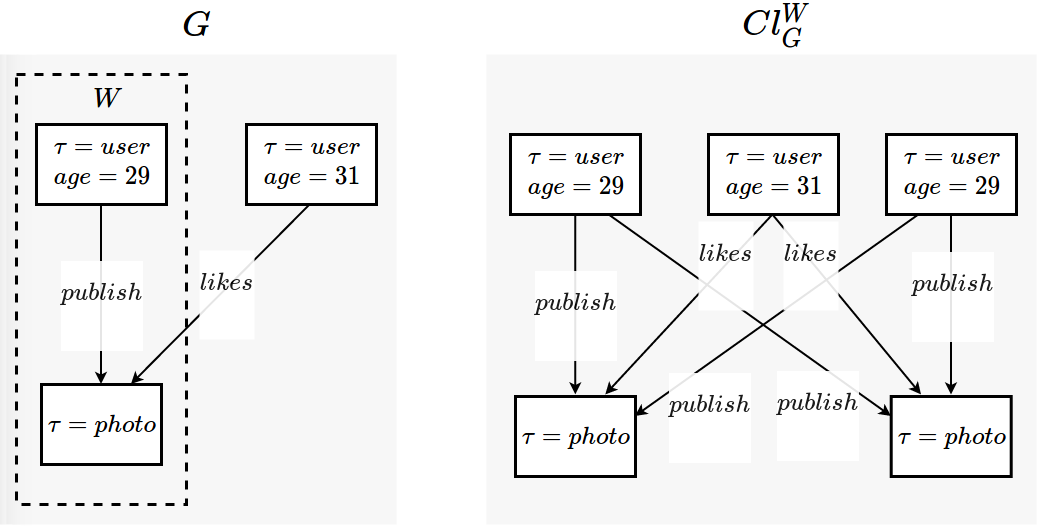}
    \end{center}
    \caption{%
        Clone of a graph
    }%
    \label{ejemplo-clonacion}
\end{figure}

Figure \ref{ejemplo-clonacion} shows an example of a cloned graph by duplicating two of its nodes. In the original graph, on the left, the set of nodes to be cloned are highlighted. The result of the cloning is presented in the graph to the right.

The following result shows that cloning positive nodes does not alter the \emph{meaning} of the queries.

\begin{theorem}
    If $Q\in GGQ$ and $W\subseteq V_Q^+$, then $Cl_Q^W\equiv Q$.
\end{theorem}
\begin{proof}
    To facilitate the notation, let $ Q_1 = Cl_Q^W $. Then, following a similar reasoning to that of the previous proof:
    \begin{align*}
    Q_1 &= \bigwedge_{n\in V_{Q_1}} {Q_1}_n^{\alpha(n)}\\
    & =\bigwedge_{n\in V_Q} {Q_1}_n^{\alpha(n)}\ \wedge \ \bigwedge_{n\in W} {Q_1}_{n'}^{\alpha(n')}\\
    &= \bigwedge_{n\in V_Q\smallsetminus \gamma_Q(W)} {Q_1}_n^{\alpha(n)}\ \wedge \ \bigwedge_{n\in\gamma_Q(W)} {Q_1}_n^{\alpha(n)}\ \wedge \ \bigwedge_{n\in W} {Q_1}_{n'}^{\alpha(n')}\\
    &= \bigwedge_{n\in V_Q\smallsetminus \gamma_Q(W)} Q_n^{\alpha(n)}\ \wedge \bigwedge_{n\in\gamma_Q(W)} Q_n^{\alpha(n)}\ \wedge \ \bigwedge_{n\in W} Q_n^{\alpha(n)}\\
    &= Q
    \end{align*}
\end{proof}

Continuing with the idea of obtaining tools to build GGQ automatically, the following concept of \emph{refinement} completes the operations that we need to refine a GGQ. A refinement set forms a kind of partition of a given GGQ.

\begin{definition}
    Given $Q\in GGQ$, $R\subseteq GGQ$ is a \emph{refinement set} of $Q$ in $G$ if:
    \begin{enumerate}
        \item $\forall\ Q'\in R\ (Q'\preceq_G Q)$
        \item $\forall\ S\subseteq G\ (S\vDash Q\Rightarrow \exists !\ Q'\in R\ (S\vDash Q'))$
    \end{enumerate}
\end{definition}

\begin{figure}[h]
    \begin{center}
        \includegraphics[scale=0.20]{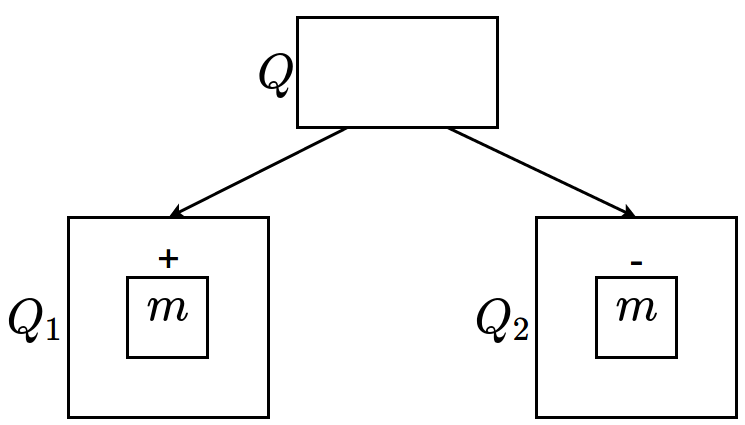}
    \end{center}
    \caption{%
        Refinement \textit{Add node}
    }%
    \label{ref1}
\end{figure}

We are now ready to present some refinement sets that will allow to automate the processes of creation and modification of Generalized Graph Query. Let's start with the simplest operation, which allows to add new nodes to an existing GGQ:

\begin{theorem}[Add new node to $Q$]
    Given $Q\in GGQ$ and $m\notin V_Q$, the set $Q+\{m\}$, formed by:
        \begin{align*}
        Q_1 &= (V_Q\cup\{m\},\ E_Q,\ \alpha_Q\cup(m,+),\ \theta_Q\cup(m,T))\\
        Q_2 &= (V_Q\cup\{m\},\ E_Q,\ \alpha_Q\cup(m,-),\ \theta_Q\cup(m,T))
        \end{align*}
    is a refinement set of $Q$ in $G$ (Fig. \ref{ref1}).
\end{theorem}

\begin{proof}
We must verify that the two necessary conditions for refinement sets are verified:
    \begin{enumerate}
        \item It is trivial that $Q\subseteq^- Q_1$ and $Q\subseteq^- Q_2$, thus $Q_1\preceq Q$ and $Q_2\preceq Q$.
        \item Given $S\subseteq G$ such that $S\vDash Q$. Then:
        \begin{align*}
        Q_1&= Q\ \wedge\ Q_m\\
        Q_2&= Q\ \wedge\ \neg Q_m
        \end{align*}
        where $Q_m=\exists v\in V\ (T)$.

        If $G\neq \emptyset$, then $S\vDash Q_1$ and $S\nvDash Q_2$.
        
        If $G= \emptyset$, then $S\nvDash Q_1$ and $S\vDash Q_2$.
    \end{enumerate}
\end{proof}

Usually, $ G \neq \emptyset $, hence this operation does not really refine, in the sense that $ Q1 \equiv Q $ and $ Q2 \equiv \neg T $. However, although we obtain an equivalent GGQ, this operation is very useful to add new nodes to a GGQ in order to add new restrictions to them later.

We proceed now to give a second refinement set that allows to create edges between existing nodes. In order to get a refinement of the original GGQ we must restrict the addition of edges to positive nodes.

\begin{figure}[h]
    \begin{center}
        \includegraphics[scale=0.30]{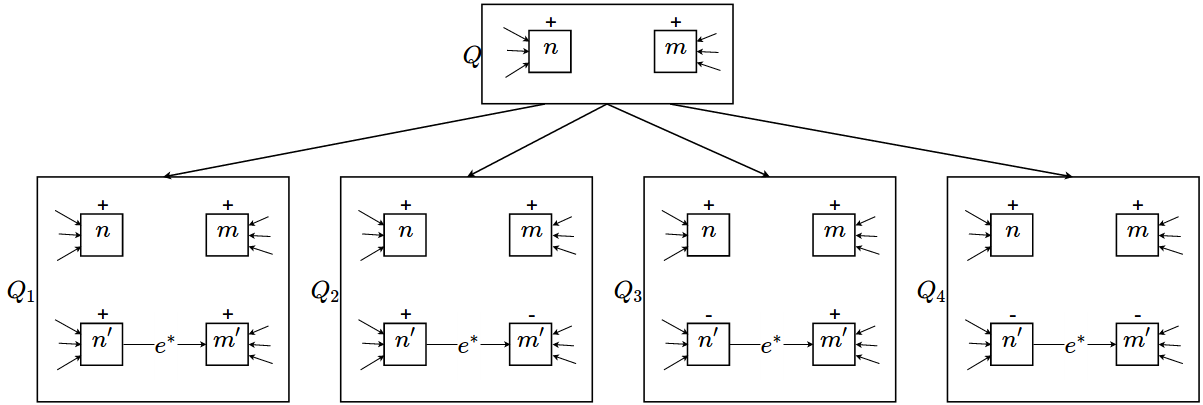}
    \end{center}
    \caption{%
        Refinement \textit{Add edge}
    }%
    \label{ref2}
\end{figure}

\begin{theorem}[Add new edge between postive nodes of $Q$]

    Given $Q\in GGQ$ and $n,m\in V_Q^+$, the set $Q+\{n^+\overset {e^*}{\longrightarrow} m^+\}$ ($*\in\{+,-\}$), formed by (where $Q'=Cl_Q^{\{n,m\}}$):
    \begin{align*}
    Q_1 &= (V_{Q'},\ E_{Q'}\cup\{n^+\overset {e^*}{\longrightarrow} m^+\},\ \theta_{Q'}\cup(e,T))\\
    Q_2 &= (V_{Q'},\ E_{Q'}\cup\{n^+\overset {e^*}{\longrightarrow} m^-\},\ \theta_{Q'}\cup(e,T))\\
    Q_3 &= (V_{Q'},\ E_{Q'}\cup\{n^-\overset {e^*}{\longrightarrow} m^+\},\ \theta_{Q'}\cup(e,T))\\
    Q_4 &= (V_{Q'},\ E_{Q'}\cup\{n^-\overset {e^*}{\longrightarrow} m^-\},\ \theta_{Q'}\cup(e,T))
\end{align*}
    is a refinement of $Q$ in $G$ (Fig. \ref{ref2}).
\end{theorem}

\begin{proof}
    \quad
    \begin{enumerate}
        \item Since $Q'$ is a clone of $Q$, and $\{n,m\}\subseteq V_Q^+$, then $Q\equiv Q'$. In addition,  $Q'\subseteq^- Q_1,Q_2,Q_3,Q_4$, thus $Q_1,Q_2,Q_3,Q_4\preceq Q'\equiv Q$.
        \item Let us consider the predicates:
        \begin{align*}
        P_n &= \exists v\in V\ \left( \bigwedge_{a\in \gamma(n)} Q_a^{\alpha(a)}\ \wedge Q_{e^o}^{\alpha(e)}\right)\\
        P_m &= \exists v\in V\ \left( \bigwedge_{a\in \gamma(m)} Q_a^{\alpha(a)}\ \wedge Q_{e^i}^{\alpha(e)}\right)
        \end{align*}
        If $S\vDash Q_n$ and $S\vDash Q_m$, then we have four mutually complementary options:
        \begin{itemize}
            \item $S\vDash P_n\ \wedge\ S\vDash P_m \Rightarrow S\vDash Q_1$
            \item $S\vDash P_n\ \wedge\ S\nvDash P_m \Rightarrow S\vDash Q_2$
            \item $S\nvDash P_n\ \wedge\ S\vDash P_m \Rightarrow S\vDash Q_3$
            \item $S\nvDash P_n\ \wedge\ S\nvDash P_m \Rightarrow S\vDash Q_4$
        \end{itemize}
    \end{enumerate}
\end{proof}
If $n=m$ ($e$ is a loop) then the previous refinement set is $\{Q_1,Q_4\}$.

Next operation adds an additional predicate to an existing edge. To keep the necessary structural conditions, this operation is restricted to positive edges connecting positive nodes.

\begin{figure}[h]
    \begin{center}
        \includegraphics[scale=0.30]{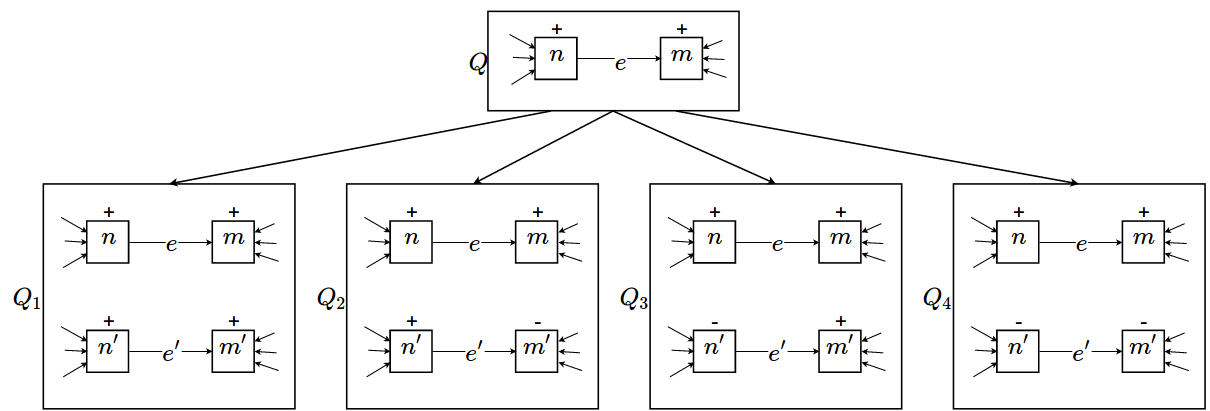}
    \end{center}
    \caption{%
	Refinement \textit{Add predicate to edge}
    }%
    \label{ref3}
\end{figure}

\begin{theorem}[Add predicate to positive edge between positive nodes of $ Q $]

    Given $Q\in GGQ$ and $n,m\in V_Q^+$, with $n^+\overset {e^+}{\longrightarrow} m^+$, and $\varphi\in FORM$, the set denoted by $Q+\{n^+\overset {e \wedge \varphi}{\longrightarrow} m^+\}$, formed by (where $Q'=Cl_Q^{\{n,m\}}$):
    \begin{align*}
    Q1 &= (V_{Q'},\ E_{Q'}\cup\{n^+\overset {e'}{\longrightarrow} m^+\},\ \theta_{Q'}\cup(e',\theta_e\wedge \varphi))\\
    Q2 &= (V_{Q'},\ E_{Q'}\cup\{n^+\overset {e'}{\longrightarrow} m^-\},\ \theta_{Q'}\cup(e',\theta_e\wedge \varphi))\\
    Q3 &= (V_{Q'},\ E_{Q'}\cup\{n^-\overset {e'}{\longrightarrow} m^+\},\ \theta_{Q'}\cup(e',\theta_e\wedge \varphi))\\
    Q4 &= (V_{Q'},\ E_{Q'}\cup\{n^-\overset {e'}{\longrightarrow} m^-\},\ \theta_{Q'}\cup(e',\theta_e\wedge \varphi))
    \end{align*}
    is a refinement set of $Q$ in $G$ (Fig. \ref{ref3}).
\end{theorem}

\begin{proof}
The proof is similar to those shown in previous results.
\end{proof}

Finally, the last operation adds predicates to existing nodes. Again, we restrict this operation to cases when the affected nodes are positive (the node where the predicate is added, and those connected to it).

\begin{figure}[h]
    \begin{center}
        \includegraphics[scale=0.35]{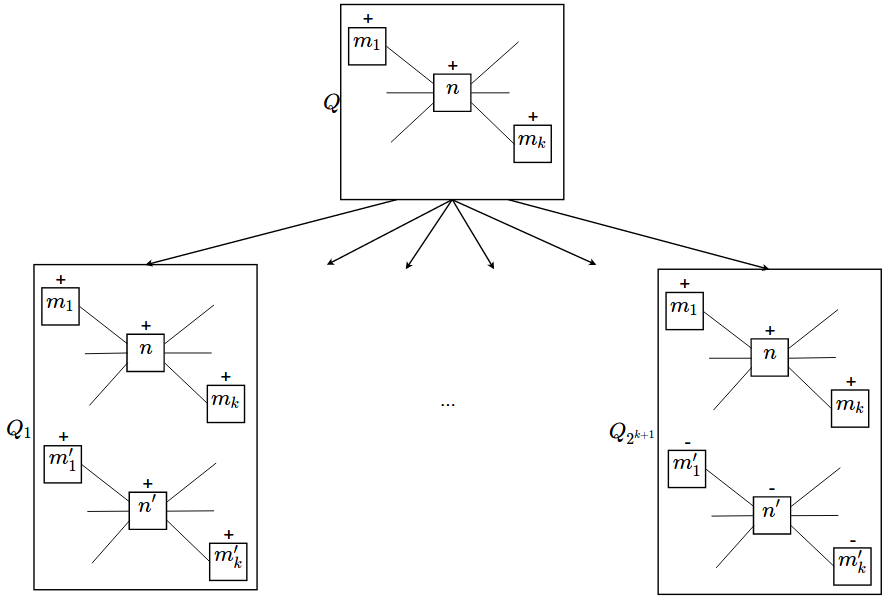}
    \end{center}
    \caption{%
        Refinement \textit{Add predicate to node}
    }%
    \label{ref4}
\end{figure}

\begin{theorem}[Add predicate to positive node with positive environment in $ Q $]

    Given $Q\in GGQ$, $n\in V_Q^+$, with $\mathcal{N}_Q(n)\subseteq V_Q^+$, and $\varphi\in FORM$. We define the set $Q+\{n\wedge \varphi\}$ formed by:
    $$\{Q_{\sigma}=(V_{Q'},E_{Q'},\alpha_{Q'}\cup \sigma,\theta_{Q'}\cup(n',\theta_n\wedge\varphi))\ :\ \sigma\in \{+,-\}^{\mathcal{N}_Q(n)}\}$$
where $Q'=Cl_Q^{\mathcal{N}_Q(n)}$, and $\{+,-\}^{\mathcal{N}_Q(n)}$ is the set of all possible assignations of signs to elements in $\mathcal{N}_Q(n)$.

Then $Q+\{n\wedge \varphi\}$ is a refinement set of $Q$ in $G$ (Fig. \ref{ref4}).
\end{theorem}

\begin{proof}
The proof is similar to the previous cases. It is only necessary to take into account that, when modifying the node $ n $, not only the $ Q $-predicate associated with it is modified but also those from all its adjacent nodes, and the set of functions $ \{+, - \}^{\mathcal{N}_Q(n)} $ cover all possible sign assignment for the nodes in the environment.
\end{proof}

It should be noted that these refinements generate structures that can be simplified. Next we provide some operations to simplify a GGQ and to obtain another equivalent and simpler one.

\begin{definition}
    Given $Q\in GGQ$, $Q'\subseteq Q$ is \emph{redundant} in $Q$ if $Q\equiv Q-Q'$. Where $Q-Q'$ is the subgraph of $Q$ given by:

    $$(V_Q\smallsetminus V_{Q'}, E_Q\smallsetminus (E_{Q'}\cup\{\gamma(n):\ n\in V_{Q'}\}),\mu_Q)$$
\end{definition}

Let us see a first result that allows to obtain simplified versions of a GGQ by removing positive redundant nodes:

\begin{theorem}
    Given $Q\in GGQ$, and $n\in V_Q^+$ such that exists $m\in V_Q$ verifying:
    \begin{itemize}
        \item $\alpha(n)=\alpha(m)$, $\theta_n\equiv\theta_m$.
        \item For each $e\in \gamma(n)$, exists $e'\in \gamma(m)$, verifying $\alpha(e)=\alpha(e')$, $\theta_e=\theta_{e'}$ and $\gamma(e)\smallsetminus\{n\}=\gamma(e')\smallsetminus\{m\}$.
    \end{itemize}
    Then, $n$ is redundant in $Q$.
\end{theorem}

Essentially, $ m $ is a clone of $ n$, but possibly with more edges connected. We can obtain a similar result for edges:

\begin{theorem}
    Given $Q\in GGQ$, and two edges, $e, e'\in E_Q$, such that $n^+\overset {e}{\longrightarrow} m^+$ and $n^+\overset {e'}{\longrightarrow} m^+$. If $\theta_e\rightarrow \theta_{e'}$ then $e'$ is redundant in $Q$.
\end{theorem}

\begin{figure}[h]
    \begin{center}
        \includegraphics[scale=0.30]{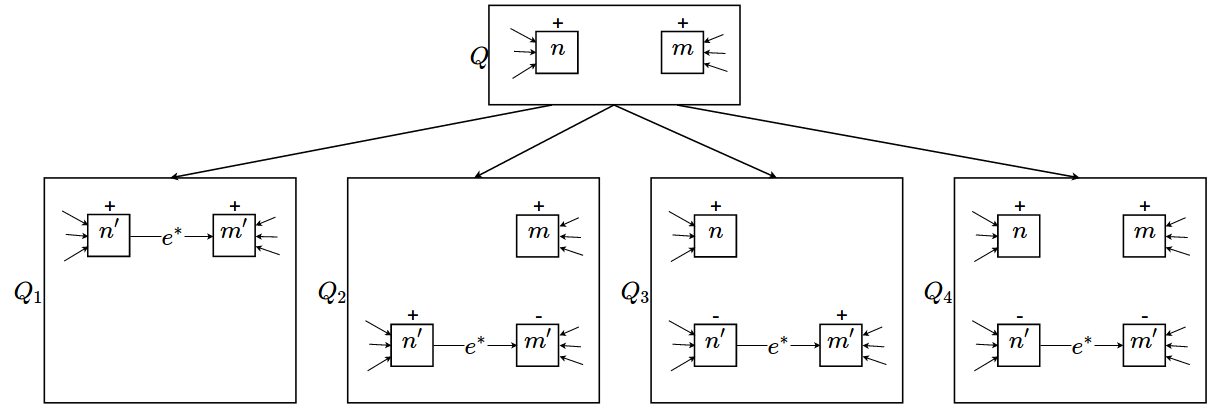}
    \end{center}
    \caption{%
        Refinement \textit{Add edge} (simplified)
    }%
    \label{ref2sim}
\end{figure}

\begin{figure}[h]
    \begin{center}
        \includegraphics[scale=0.30]{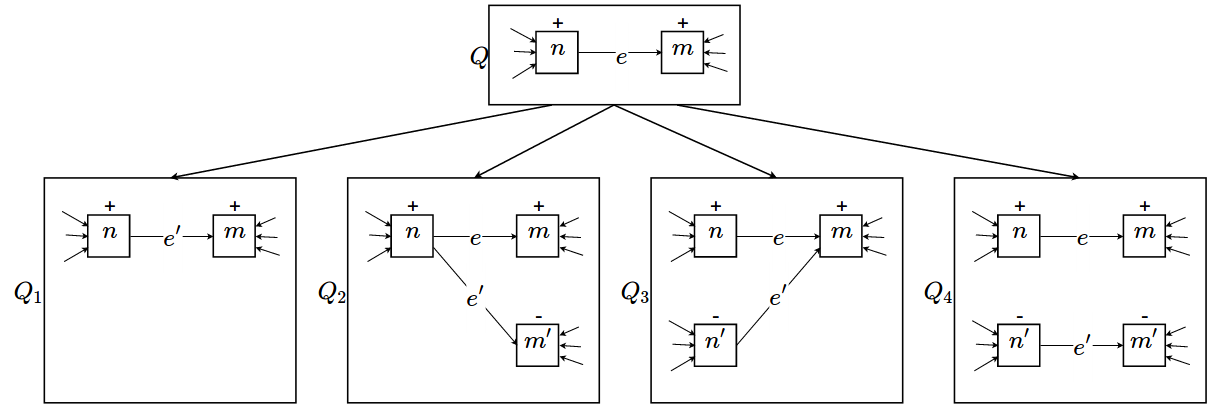}
    \end{center}
    \caption{%
        Refinement \textit{Add predicate to edge} (simplified)
    }%
    \label{ref3sim}
\end{figure}

\begin{figure}[h]
    \begin{center}
        \includegraphics[scale=0.30]{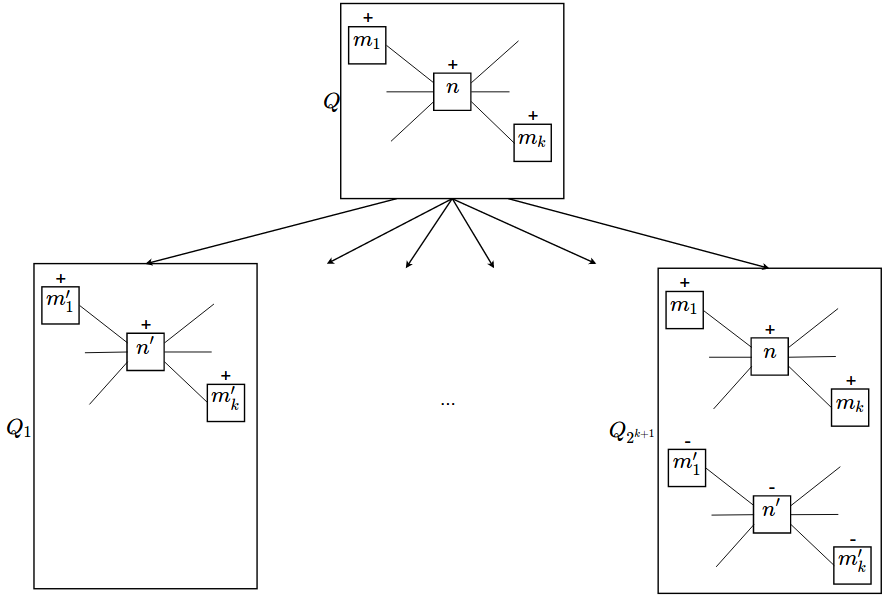}
    \end{center}
    \caption{%
        Refinement \textit{Add predicate to node} (simplified)
    }%
    \label{ref4sim}
\end{figure}

From these two results we can give simplified versions of the previous refinement sets, grouping positive nodes and positive edges when, after initial cloning, the sign of the duplicate element has been maintained with the original, as well as in the cases where the sign has been maintained and an additional predicate has been added. Figures \ref{ref2sim} to \ref{ref4sim} shows representations of the refinement sets $Q+\{n\wedge \varphi\}$, $Q+\{n^+\overset {e \wedge \varphi}{\longrightarrow} m^+\}$ and $Q+\{n\wedge \varphi\}$ applying these simplifications.

\begin{figure}[h]
    \begin{center}
        \includegraphics[scale=0.35]{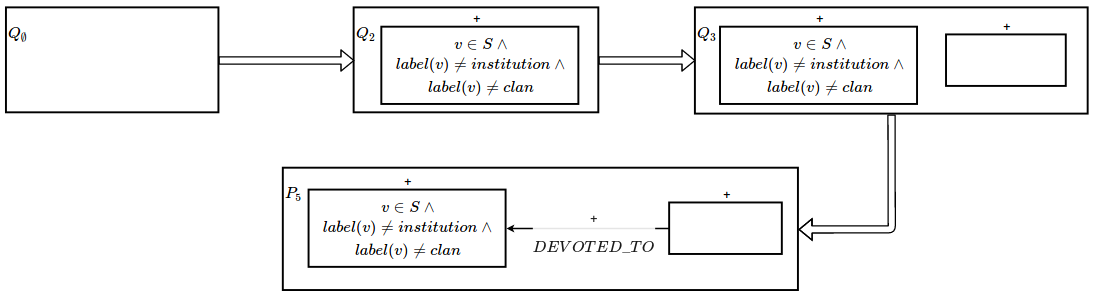}
    \end{center}
    \caption{%
		Sequences of refinements for $ P_5 $
    }%
    \label{pqg5refs}
\end{figure}

For example, the following sequence of refinements constructs the pattern $ P_5 $ (Fig. \ref{pqg5refs}):
\begin{align*}
Q_1 &= Q_\emptyset+\{n_1\}\\
Q_2 &= Q_1+\{n_1 \land (v \in S \land \tau(v) \neq \mathtt{institution} \land \tau(v) \neq \mathtt{clan} \}\\
Q_3 &= Q_2+\{n_2\}\\
Q_4 &= Q_3+\{n_2 \overset{e_1}{\longrightarrow} n_1\}\\
P_5 &= Q_4+\{n_2 \overset{e_1 \land (\tau(\rho) = \mathtt{DEVOTED\_TO})}{\longrightarrow}n_1\}
\end{align*}

\begin{figure}[h]
    \begin{center}
        \includegraphics[scale=0.25]{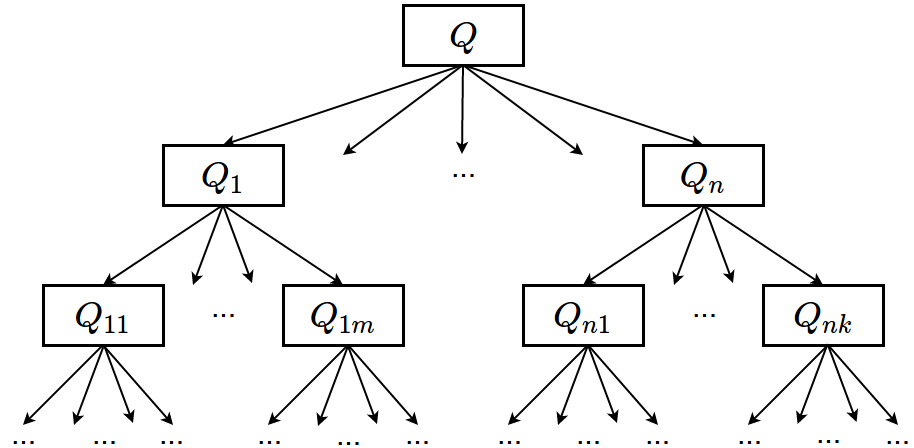}
    \end{center}
    \caption{%
        Refinements tree
    }%
    \label{arbolGGQ}
\end{figure}

From the structure of a GGQ it is not easy to obtain a complementary GGQ with it. However, there are many analysis on property graphs (or generalized graphs) where we need to work with sequences of queries verifying some properties of containment and complementarity as predicates. The refinements presented in this section come to cover this gap and to allow, for example, the construction of an embedded partition tree  with the nodes labelled as follows (Fig. \ref{arbolGGQ}):

\begin{itemize}
    \item The root node is labelled with $ Q_0 $ (some initial GGQ).
    \item If a tree node is labelled with $ Q $, and $ R = (Q_1, \dots, Q_n) $ is a refinement set of $ Q $, then its child nodes are labelled with the elements of $R$.
\end{itemize}

Note that the construction of this tree completely depends on the refinement chosen in each branch, and the initial GGQ.

The refinements presented here are only one option, but not the only one. For example, we could consider refinements that, instead of adding constraints to positive elements, lighten the conditions over negative elements, and using disjunction of predicates instead of conjunction of them.

\section{Conclusions and Future Work}

In this work we have presented a framework to evaluate subgraphs immersed in property graphs (more generally, in generalized graphs) that can be used in discovery procedures over relational data. We want this framework to verify several requirements:
\begin{itemize}
	\item To use the same grammar for the queries and for the structures to evaluate. Thanks to the expressive power of generalized graphs we have presented a query tool that can be expressed by using generalized graphs.
	\item To provide well-founded basis that would assure the queries behave consistently and robustly. These results have been obtained by studying the relationships between the topological structure of the query and the logical meaning of the query.
	\item In addition, we have provided a controlled way to construct GGQ by means of atomic operators that translates the topological control of the construction into a logical control of the meaning. In this sense, we have introduced a first family of refinements to achieve this goal.
\end{itemize}

Because relational data can be viewed as graphs, and queries can be viewed as pattern searchs, most query languages in databases can be viewed as (perhaps primitive) graph pattern matching tools. In this paper we have analysed some of the existing query tools as well as the feasibility to be used in automatic procedures. One of these tools, Selection Graphs, allows to evaluate records in relational databases through acyclic patterns that can be refined by using basic operations, and allowing to obtain complementary patterns in each case. They do not require an exact projection of the pattern representing the selection graph onto the subgraph to be evaluated, but rather the fulfillment of a series of predicates expressed in the pattern. We must remember that if a projection is required when carrying out the verification of a pattern, the task of evaluating the non-existence of certain elements becomes hard. Specifically, selection graphs evaluate the existence / non-existence of paths incidents into the record under evaluation (they are only capable of evaluating individual records) by verifying a conjunction of predicates associated to those paths, and it can be seen as the evaluation of existence of a tree rooted in the node that represents the record under evaluation.

Generalized Graph Query extends the concept of selection graphs allowing the evaluation of general subgraphs, beyond a single node, the use of more powerful predicates and allowing cyclical patterns. As it becomes a requirement not to use a projection for the verification of a pattern, these objectives have been achieved by extending the form of evaluation, which can be seen as the evaluation of a tree rooted in every node from the pattern (allowing the edges to be identified with paths in the graph). Consequently, it manages more complex structures. The intersections between the various trees and the constraints imposed on the nodes allow the evaluation of cyclical patterns in the GGQ, something that had not been achieved in previous proposals.

As we have mentioned, like selection graphs, GGQ can be modified and constructed from refinements, but unlike the simple case of selection graphs, refinements of GGQ are usually not binary, resulting in sets of size $ 2 ^ k $ (where $ k $ is the number of modified predicates). In general, refinements result in embedded partitions of the structures they evaluate, making them ideal tools for white-box machines learning procedures. After carrying out a first (but fully functional) proof-of-concept implementation, it has been experimentally demonstrated the practical usefulness of GGQ framework.

An explicit use of these capabilities has already been carried out in knowledge discovery procedures, specifically in the GGQ-ID3 algorithm, which makes use of Generalized Graph Query as tests in the nodes of decision trees. The relationship between GGQ and GGQ-ID3 is equivalent to that between selection graphs and the algorithm MRDTL \cite{Leiva02mrdtl:a}. In the experiments carried out in this context it is shown that GGQ-ID3 is able to extract interesting GGQ patterns that can be used in complex tasks.

On the other hand, more complex refinement families can be created (for example, combining the refinement \emph{add edge} with \emph{adding property to an edge} in a single step) to thereby reduce the number of steps to get complex GGQ and to get more powerful atomic steps. If this option is carried out properly (unifying the refinements according to the frequency of occurrence of structures in a graph, for example) faster version of algorithms making use of GGQ can be obtained. In this case the improvement in efficiency is achieved by sacrificing the possibility of covering a wider query space. A minimal and robust set of refinement operations have been offered in this work, but they are not intended to be optimal for every learning task. Consequently, GGQ represents a powerful and simple query tool of controlled complexity, suitable for automatic construction and to be used in white-box multi-relational machine learning algorithms.

In the future work that derives from the development presented here, it is worth mentioning that, since GGQ are constructed using the generalized graph structure (that allows the definition of hypergraphs in a natural way), with slightly modifications GGQ can evaluate hypergraphs with properties. Hence, the extension of Generalized Graph Query to Generlized Hypergraph Query is a natural step that is worth considering. The development of different refinement sets in correspondence to the learning task or the type of graph to query is a future line of research. In addition, the automatic generation of such sets from statistics extracted from the graph to be analysed can lead to important optimizations in GGQ automatic construction processes. Finally, it should be noted that GGQ is already being used by discovery / learning procedures, such as the multi-relational tree-building algorithm GGQ-ID3 named above, and thanks to its good properties, it is a great candidate to be used by other algorithms of this type.

\bibliographystyle{plain}	% or "siam", or "alpha", etc.
\bibliography{biblio}

\end{document}

\title{Consulta de Patrones en Grafos Generalizados}
\author{Pedro Almagro-Blanco and Fernando Sancho-Caparrini}
\maketitle

\section{Introducción}

Debido al rápido crecimiento de las tecnologías relacionadas con Internet, y a que los grafos poseen una alta capacidad expresiva y son especialmente adecuados para modelar estructuras de elevada complejidad \cite{fast-graph-pattern}, el número de aplicaciones que hacen uso de ellos para modelar tanto sus datos como los procesos que los manipulan ha crecido espectacularmente en los últimos años.

El uso de nuevas estructuras conceptuales en las aplicaciones del \textit{mundo real} requiere del desarrollo de nuevos sistemas para almacenar y consultar dichas estructuras, de tal forma que los usuarios puedan acceder a los datos almacenados de manera efectiva y eficiente. Como ocurre con toda nueva tecnología que se implementa, a pesar de que está demostrando una gran madurez y puede dar respuesta a muchas de las necesidades de los usuarios, todavía se encuentra en fase de desarrollo y en búsqueda de un conjunto de estándares que aseguren un crecimiento continuado y sin sobresaltos.

La creciente popularidad de las Bases de Datos en Grafo ha dado lugar a la aparición de interesantes problemas relacionados con el almacenamiento y consulta en este tipo de soluciones. Haciendo un paralelismo con la evolución tecnológica que vivieron los modelos de datos relacionales, estas bases de datos disponen ya de unos fundamentos robustos en lo referente a los accesos básicos de la información (creación, acceso, eliminación y modificación de elementos individuales de la estructura), pero todavía adolecen de estándares robustos en algunas de las otras tareas necesarias para el almacenamiento y recuperación de la información, donde destacan principalmente las tareas relacionadas con mecanismos más avanzados de consulta.

Entre los problemas relacionados con estos procesos de consulta, la detección de patrones está considerada como uno de los problemas fundamentales ya que engloba otros subproblemas necesarios para la obtención de sistemas de consulta potentes, como son la búsqueda de subgrafos, la búsqueda de caminos mínimos, o el estudio de la conectividad \cite{distance-join,gupta2015neo4j}.

En este artículo presentaremos Generalized Graph Query (GGQ), una propuesta para llevar a cabo consultas de patrones en grafos con propiedades. Los GGQ presentan un marco lógico robusto, por lo que son especialmente útiles en procedimientos de descubrimiento en grafos y generalizan otras herramientas más básicas que han demostrado ser muy útiles en tareas de este tipo. Además, presentaremos una colección de operaciones de refinamiento que permite construir GGQs a partir de un GGQ inicial, y que será muy útil a la hora de automatizar la construcción automática de este tipo de consultas.

Este artículo se estructura como sigue. Comenzaremos haciendo un análisis del estado del arte de \textit{Graph Pattern Matching}, recorriendo los conceptos fundamentales en este área y analizando las herramientas más importantes que permiten realizar consultas de patrones en grafos, que culminará con la presentación de nuestra propuesta, los Generalized Graph Query, presentando algunas de sus propiedades generales y una colección de ejemplos que nos acercarán a posibles usos más elaborados. A continuación mostraremos una posible familia de operaciones sobre GGQ que permiten realizar construcciones de consultas de forma controlada. Por último, presentamos algunas de las conclusiones obtenidas durante la realización de este trabajo así como algunas de las líneas de trabajo futuro que han derivado del mismo.

\section{Graph Pattern Matching}

La detección de patrones en grafos (\emph{Graph Pattern Matching}) es un área de investigación activa desde hace más de 30 años que ha mostrado su utilidad en muy diversas áreas del conocimiento, desde la visión artificial, hasta la biología, pasando por la electrónica, el diseño asistido por ordenador, y el análisis de redes sociales, entre otros. Sin duda, su interés ha crecido aún más al hacer su aparición las bases de datos en grafo como una herramienta de estructuración y almacenamiento de la información de forma transversal común a todas las áreas de conocimiento. Por este motivo, el problema de la detección de patrones en grafos se expande, con ligeras variantes, a través de diferentes comunidades científicas, mostrándose no como un problema único definido bajo una formalización común, sino como un conjunto de problemas relacionados.

El proceso por el cual comprobamos la presencia de un determinado patrón en un conjunto concreto de datos se denomina {\it detección de patrones}, y computacionalmente comprende el conjunto de procesos mecánicos que permiten dar como respuesta una (o todas) las \emph{ocurrencias} del patrón.

La definición exacta de qué se entiende por una \emph{ocurrencia} de un patrón en un grafo varía según cómo se defina el patrón. Cuando éste viene dado por una estructura de grafo (aunque no necesariamente usando los mismos conjuntos elementales de vértices y aristas que el grafo sobre el que se realiza la consulta) es habitual asociar esta ocurrencia con la existencia de identificaciones (quizás no tan fuertes como isomorfismos) entre el patrón y aquellos subgrafos del grafo de datos que respeten algunas restricciones impuestas por patrón \cite{gpmsocial}.

Algunas clasificaciones rápidas que podemos presentar en cuanto a las diferentes maneras posibles de llevar a cabo la detección de patrones en grafos son: (a) \emph{Detección estructural vs. Detección semántica}, (b) \emph{Detección exacta vs. Detección inexacta.}, y (c) \emph{Solución óptima vs. Solución aproximada.} \cite{matching}.

Junto a estas clasificaciones se puede dar otra atendiendo a la relación que debe cumplir el patrón con respecto a los subgrafos que se consideren ocurrencias del mismo y que permite dividir las diferentes técnicas de detección de patrones en grafos en aquellas basadas en \emph{Isomorfismos de Subgrafos} \cite{ullmann} (donde una ocurrencia de un patrón en un grafo será cualquier subgrafo que sea isomorfo al patrón), \emph{Graph Simulation} \cite{Milner} (donde una ocurrencia será cualquier subgrafo para el que existe una relación binaria entre los elementos del subgrafo y el patrón, que respete los tipos de los nodos y las adyacencias de los mismos), \emph{Bounded Simulation} \cite{Fan,distance-join, Milner} (basado en Graph Simulation pero permitiendo asociar aristas del patrón a caminos del subgrafo) y \emph{Regular Pattern Matching} \cite{rexp,phdthesis,Barcelo, Milner} (basado en Bounded Simulation pero permitiendo restringir los caminos con los que se identifican las aristas del patrón a través de expresiones regulares). 

Las herramientas actuales que permiten realizar consultas de patrones en grafos utilizan, o bien un lenguaje declarativo (como Cypher, SPARQL, o SQL), o bien un lenguaje imperativo (con Gremlin como más claro representante). En el caso de los lenguajes declarativos es responsabilidad del sistema (idealmente) llevar a cabo la optimización automática de las consultas, preocupándose el usuario únicamente de declarar el objeto de su consulta. En el caso de las aproximaciones imperativas el plan de ejecución es responsabilidad del usuario (que debe tener conocimientos de desarrollo), por lo que suelen proporcionar aproximaciones a más bajo nivel.

A continuación presentamos, brevemente, algunos de los lenguajes de consulta que se han utilizado para la detección de patrones en grafos. 

\emph{SQL} (Structured Query Language), es un lenguaje declarativo para acceder a Sistemas de Gestión de Bases de Datos Relacionales (RDBMS). Este tipo de bases de datos fueron diseñadas para datos tabulares con esquema fijo, y trabajan mejor en contextos que están bien definidos desde el principio y en los que los propios registros son más importante que las relaciones entre ellos (las consultas se llevan a cabo habitualmente imponiendo restricciones sobre propiedades de los registros y no sobre las relaciones en las que participan). Por ello, intentar responder a preguntas en las que se involucran muchas relaciones entre los datos (como suele ser habitual en  las consultas de patrones en grafos) con una base de datos relacional implica numerosas y costosas operaciones entre las tablas que pueden llegar a hacer inviable esta opción. A pesar de ello, SQL posee la capacidad expresiva necesaria para poder expresar las consultas de patrones en grafos más habituales. Por ello, y porque las bases de datos relacionales han sido la opción de almacenamiento elegida por la mayoría de los proyectos durante décadas y SQL es su lenguaje de consulta por excelencia, los primeros sistemas de consulta de patrones en grafos también usaron consultas SQL, como es el caso de los \textit{Grafos de Selección} que veremos más adelante.

\emph{SPARQL} es un lenguaje de consulta declarativo para atacar datos almacenados en formato RDF \cite{Segaran:2009:PSW:1696488} y es reconocido como una de las tecnologías claves de la Web Semántica, por ello, su capacidad expresiva para consultas de patrones en grafos es muy superior a la que proporciona SQL, pero queda lejos de ser intuitiva para un usuario humano. SPARQL permite consultas de patrones estructurales, semánticos, óptimos, y exactos basados en isomorfismos de subgrafos. A pesar de que SPARQL no permite realizar ningún tipo de Graph Simulation ni Regular Pattern Matching se han desarrollado extensiones del lenguaje como PSPARQL \cite{alkhateeb2007rdf} que permiten consultar bases de datos RDF utilizando patrones que hacen uso de expresiones regulares. Además, gracias a la estructura que presenta el lenguaje, es relativamente fácil la generación automática de consultas en SPARQL, por lo que existen otras herramientas que lo utilizan como lenguaje de consulta final.

\emph{Gremlin} es, simultáneamente, un lenguaje de consultas para bases de datos en grafo y una máquina virtual orientada a realizar computación sobre grafos. Como lenguaje, Gremlin es independiente de la base de datos utilizada (por medio de conectores). Su sintaxis recuerda a la programación funcional, permitiendo realizar consultas declarativas e imperativas, ya que está basado en un flujo de datos que permite a los usuarios expresar de manera sencilla consultas complejas en grafos. De esta forma, cada consulta está compuesta por una secuencia de pasos que realizan operaciones atómicas en el flujo de datos. Gremlin permite realizar consultas semánticas, exactas, óptimas, y basadas en isomorfismos de subgrafos de manera natural. Dado que Gremlin es un lenguaje, un juego de instrucciones y una máquina virtual, es posible diseñar otros lenguajes de consulta en grafos que compilen al lenguaje Gremlin (por ejemplo, SPARQL puede ser compilado para ejecutarse en una máquina Gremlin\footnote{\url{https://github.com/dkuppitz/sparql-gremlin}}).

Los \emph{Grafos de Selección} son un tipo de patrones multi-relacionales para consultar bases de datos basadas en la tecnología SQL. Representan los cimientos sobre los que está construido el algoritmo de inducción de árboles de decisión multi-relacionales MRDTL \cite{Leiva02mrdtl:a} así como otros algoritmos de aprendizaje automático multi-relacionales \cite{DBLP:journals/corr/abs-1211-3871}. El objetivo final de los grafos de selección es su uso en procedimientos de búsqueda de patrones de tipo \textit{top-down} \cite{Knobbe99multi-relationaldecision}, y para ello se necesitan operadores que permitan modificar, a través de pequeños cambios, un grafo de selección dado. Además, permiten una representación gráfica muy expresiva y pueden ser construidos en pasos sucesivos, ofreciendo buenas condiciones para ser utilizados en procedimientos de descubrimiento. Es por estas razones por las que nuestra propuesta se puede considerar una generalización de algunas de las ideas que promovieron esta aproximación. Como contraparte, presentan el inconveniente de que los patrones que representan no pueden contener ciclos, limitando así la potencia expresiva de las consultas. Los grafos de selección representan un tipo de Graph Pattern Matching exacto, óptimo y basado en el isomorfismo semántico de grafos. Además, al ser una representación gráfica de las consultas SQL, hereda los problemas de eficiencia que presentan los sistemas basados en esta tecnología.

\emph{Cypher} es un lenguaje de consulta declarativo desarrollado específicamente para trabajar sobre la base de datos en grafo Neo4j\footnote{\url{http://neo4j.org}}.  Cypher está diseñado para ser un lenguaje de consulta \textit{humano}, cercano tanto para desarrolladores como para usuarios finales, y las consultas de patrones en grafos que habitualmente son complicadas en otros lenguajes resultan muy sencillas en él \cite{cypher}, mostrando una alta capacidad expresiva \cite{cypherpat}. La base de datos Neo4j sigue con mucha fidelidad el modelo de grafo con propiedades, pero obliga que las aristas tengan un tipo asociado. A diferencia con Gremlin, Cypher no es Turing completo, por lo que presenta algunas limitaciones (por ejemplo, no es capaz de llevar a cabo algunos algoritmos de análisis en grafos), y es de más alto nivel (por ejemplo, no es capaz de expresar la forma en la que se quiere paralelizar una consulta). Las consultas sencillas en Cypher poseen un buen rendimiento, sin embargo no siempre es así cuando las consultas siguen patrones complejos, ya que cuando hay condiciones múltiples Cypher no permite indicar en qué orden aplicar dichas condiciones. Cypher permite consultas de patrones en grafos estructurales y semánticas, óptimas, exactas, y basadas en el isomorfismo de subgrafos. Además, permite un tipo de Regular Pattern Matching en el que las aristas en el patrón se proyecten sobre caminos del grafo, y se pueden imponer restricciones a esos caminos a través de expresiones que hacen uso del operador disyunción y del cierre de Kleene. Una limitación desde el punto de vista académico es que carece de un modelo formal asociado, y se ha construido con un carácter completamente aplicado, por lo que algunas de sus operaciones no han sido validadas. A pesar de ello, por su excelente expresividad y aceptable rendimiento, y porque consume los datos de una base de datos en grafo muy extendida en su uso, Cypher es el lenguaje base que se ha elegido para implementar Generalized Graph Query, nuestra propuesta para llevar a cabo consultas de patrones en grafos con propiedades.

Algunas otras herramientas relacionadas con la consulta de patrones en grafos son: \emph{GraphLog} \cite{graphlog}, que permite estructurar las consultas en forma de grafo y evalúa la existencia de un patrón determinado entre un par de nodos para generar una nueva arista entre éstos; \emph{GraphQL}\footnote{\url{http://graphql.org/}}, lenguaje de consulta declarativo desarrollado por la compañía Facebook para permitir el acceso a su información por parte de aplicaciones externas; \emph{Graql}\footnote{\url{https://grakn.ai}}, lenguaje de consulta declarativo orientado a grafos de conocimiento; y \emph{PGQL} \cite{van2016pgql}, que representa una extensión SQL con características propias de las consultas en grafos: análisis de accesibilidad entre nodos, localización de caminos, y construcción de grafos. 

\section{Definiciones Previas}

Dado $V$ un conjunto cualquiera, denotaremos por:
$$V^0= \emptyset,\ V^1= V,\ V^{n+1} = V^n \times V,\ V^*=\bigcup_{n\geq 0}V^n$$

En general, a los elementos de $V^*$ los llamaremos \emph{secuencias}, \emph{sucesiones} o \emph{listas}. Si $x\in V^n$ entonces diremos que $x$ tiene longitud $n$, y escribiremos $|x|=n$.

Si $x=(a_1,\dots,a_n),\ y=(b_1,\dots,b_m)\in V^*$, entonces la \emph{concatenación} de $x$ e $y$ es el elemento de $V^*$ dado por $xy=(a_1,\dots,a_n,b_1,\dots,b_m)$.

Para cada $x=(a_1,\dots,a_n)\in V^n$, llamaremos \emph{conjunto soporte de $x$} al conjunto $s(x)=\{a_i:\ 1\leq i\leq n\},$.
y, por un abuso del lenguaje, escribiremos $a\in x$ para indicar que $a\in s(x)$.

Para cada $a\in V$, denotamos $|a|_x=\# \{i:\ x_i=a\}$ (donde $\#(A)$ denota el cardinal del conjunto $A$), y llamaremos \emph{multiconjunto soporte de $x$} al conjunto de pares $ms(x)=\{(a,|a|_x):\ a\in x\}$.

A partir de los multiconjuntos soporte podemos definir la relación $\sim$, que se puede probar fácilmente que es de equivalencia en $V^n$, como: $x \sim y$ si y solo si $ms(x)=ms(y)$, 
y denotaremos por $V_\sim^n=V^n/\sim$ (conjunto cociente de $V^n$ bajo la relación $\sim$).

Nuestra interpretación de estos conjuntos será que, así como $V^n$ denota el conjunto de tuplas ordenadas de elementos de $V$, $V_\sim^n$ denota el conjunto de tuplas no ordenadas del mismo conjunto (es decir, tuplas en las que importan los elementos que aparecen, considerando las posibles repeticiones, pero no el orden en el que aparecen).

A continuación presentamos la definición de \emph{Grafo Generalizado}, que abarca las diferentes variantes de grafo que se pueden encontrar en la literatura y que necesitaremos a la hora de presentar nuestra propuesta de consulta de patrones en grafos.

\begin{definition}
Un \emph{Grafo Generalizado} es una tupla $G = (V,E,\mu)$ donde:
\begin{itemize}
	\item $V$ y $E$ son conjuntos, que llamaremos, respectivamente, \emph{conjunto de nodos} y \emph{conjunto de aristas} de $G$.
	\item $\mu$ es una relación (habitualmente la consideraremos funcional, pero no es necesario) que asocia a cada nodo o arista en el grafo su conjunto de propiedades, es decir, $\mu : ( V \cup E) \times R \rightarrow S$, donde $R$ representa el conjunto de posibles \emph{claves} para dichas propiedades, y $S$ el conjunto de posibles \emph{valores} asociados a las mismas.
\end{itemize} 

Habitualmente, para cada $\alpha \in R$ y $x\in V\cup E$, escribiremos $\alpha(x)=\mu(x,\alpha)$.

Además, exigiremos la existencia de una clave destacada para las aristas del grafo, que llamaremos \emph{incidencias} y denotaremos por $\gamma$, que asocia a cada arista del grafo una tupla, ordenada o no, de vértices del grafo.
\end{definition}

Aunque la definición que hemos presentado aquí es más general que las que se pueden encontrar en la literatura relacionada, también los denominaremos \emph{Grafos con Propiedades}, ya que suponen una extensión natural de este tipo de grafos. 

Cabe indicar que en los grafos generalizados que acabamos de mostrar, y a diferencia de las definiciones tradicionales, los elementos en $E$ son símbolos que representan a las aristas y no pares de elementos de $V$, y es $\gamma$ la función que asocia a cada arista el conjunto de vértices que relaciona. 

\begin{definition}[Notación y definiciones]
	En el contexto de las definiciones anteriores, usaremos la siguiente notación:
\begin{itemize}
\item Habitualmente identificaremos $e$ con $\gamma(e)$, de forma que si $v\in V$ escribiremos $v\in e$ para denotar que $v\in \gamma(e)$. Así, interpretamos la arista como la colección de nodos que conecta, tal y como siguen las definiciones más clásicas de grafos.
\item De forma simétrica, para cada $u\in V$ escribiremos $\gamma(u)=\{e\in E:\ u\in e\}$.
\item En general, un grafo generalizado puede tener combinación de aristas dirigidas y no dirigidas. Si $\gamma:E\rightarrow V^*$ diremos que el grafo es \emph{Dirigido}. Si $\gamma:E\rightarrow V_\sim^*$ diremos que el grafo es \emph{No Dirigido}. 
\item Para cada $e\in E$, se define la \emph{aridad} de $e$ como $\sum_{a\in e} |a|_e$.
\item Si $\gamma:E\rightarrow V^2 \cup V_\sim^2$ diremos que el grafo es \emph{Binario} (y coincide con la estructura de grafo más habitual). En caso contrario, diremos que el grafo es un \emph{Hipergrafo}.
\item Una arista, $e\in E$, se dice que es un \emph{lazo} si conecta un nodo con él mismo, es decir, si tiene aridad distinta a 1 pero $s(e)$ es unitario.
\item Una arista, $e\in E$, se dice \emph{incidente} en un nodo, $v\in V$, si $v\in e$.
\item Dos nodos distintos, $u,v\in V$ se dicen \emph{adyacentes}, o \emph{vecinos}, en $G$ si existe $e\in E$ tal que $\{u,v\}\subseteq e$.
\item Si existen aristas distintas en $E$ con la misma incidencia, es decir, aristas que conectan los mismos nodos, diremos que el grafo es un \emph{Multi-grafo}.
\item Si $e$ es una arista binaria dirigida que conecta $u$ con $v$, $e=(u,v)$, escribiremos $u\overset{e}{\rightarrow}v$, y también notaremos $e^o=u$ (\emph{output} de $e$) y $e^i=v$ (\emph{input} de $e$). En este caso, para cada $u\in V$ escribiremos: $$\gamma^o(u)=\{e\in \gamma(u):\ e^o= u\}$$ $$\gamma^i(u)=\{e\in \gamma(u):\ e^i=u\}$$
que denotan, respectivamente, el conjunto de \emph{aristas salientes} de $u$ y el conjunto de \emph{aristas entrantes} en $u$.
\item Dado $u\in V$, definimos el \emph{entorno} de $u$ en $G$ como el conjunto de nodos, incluyendo a $u$, que están conectados con él, es decir: $\mathcal{N}(u)=\bigcup_{e\in \gamma(u)}\gamma(e)$.
Cuando sea necesario hablaremos del \emph{entorno reducido} de $u$ como $\mathcal{N}^*(u)=\mathcal{N}(u)\setminus \{u\}$.
\end{itemize}
\end{definition}

La noción de subgrafo se obtiene de la definición habitual añadiendo a las condiciones habituales de contención de nodos y aristas la condición de que las propiedades también se mantengan en los elementos comunes.

\begin{definition}
Un subgrafo de un grafo $G = (V, E,\mu)$ es un grafo $S = (V_S, E_S,\mu_S)$ tal que $V_S \subseteq V$ y $E_S \subseteq E$ y $\mu_S\subseteq \mu_{|V_S \cup E_S}$. Notaremos $S \subseteq G$. 
\end{definition}

Un concepto fundamental al trabajar con grafos es el de \emph{camino}, que permite estudiar relaciones de distancia y condiciones de conectividad entre diferentes elementos, extendiendo la conectividad de las aristas a situaciones más generales.

Debido a que nuestros grafos son considerablemente más generales que los habituales (hasta el punto de contener el concepto de hipergrafo, que generalmente no se cubre en la Teoría de Grafos clásica) hemos de dar previamente algunas nociones que permitan hablar de la posición de orden que ocupa un nodo en una arista:

\begin{definition}
Si $e\in E$ y $\gamma(e)=(v_1,\dots,v_n)\in V^n$, entonces para cada $v_i\in s(e)$ definimos su \emph{orden} en $e$ como $ord_e(v_i)=i$. Si $e\in V_\sim^n$, entonces para cada $v\in s(e)$ definimos $ord_e(v)=0$.

Este orden define de forma natural un orden entre los nodos incidentes en una arista, y escribiremos $u\leq_e v$ para indicar que $ord_e(u)\leq ord_e(v)$.
\end{definition}

A partir de esta relación de orden entre los nodos que conecta una arista, podemos definir de manera general qué entendemos por un camino dentro de un grafo.

\begin{definition}
Dado un grafo $G=(V,E,\mu)$, el conjunto de \emph{caminos} en $G$, que denotaremos por $\mathcal{P}_G$, se define como el menor conjunto verificando las siguientes condiciones:
\begin{enumerate}
	\item Si $e\in E$, $u,\ v\in e$ con $u \leq_e v$, entonces $\rho=u\overset{e}{\rightarrow}v\in \mathcal{P}_G$, y $sop_V(\rho)=(u,v)$, $sop_E(\rho)=(e)$. Diremos que $\rho$ \emph{une} (o \emph{conecta}) los vértices $u$ y $v$ de $G$, o que \emph{$v$ es accesible desde $u$ por medio de $\rho$}, y lo notaremos por $\conecta{u}{\rho}{v}$.
	\item Si $\rho_1,\ \rho_2\in \mathcal{P}_G$, con $\conecta{u}{\rho_1}{v},\ \conecta{v}{\rho_2}{w}$, entonces $\rho_1\cdot\rho_2 \in \mathcal{P}_G$, con $\conecta{u}{\rho_1\cdot\rho_2}{w}$, $sop_V(\rho_1\cdot\rho_2)=sop_V(\rho_1)sop_V(\rho_2)$, $sop_E(\rho_1 \cdot \rho_2)=sop_E(\rho_1)sop_E(\rho_2)$.
\end{enumerate}
En caso de que $u=v$ diremos que $\rho$ es un camino \emph{cerrado}, y si además no se repiten aristas en $\rho$ diremos que es un \emph{ciclo}.
\end{definition}

Si $\rho\in\mathcal{P}_G$, con $sop_V(\rho)=(u_1,\dots,u_{n+1})$ y $sop_E(\rho)=(e_1\dots,e_n)$, entonces escribiremos:
$$\rho=u_1 \overset{e_1}{\rightarrow} u_2 \overset{e_2}{\rightarrow} \dots \overset{e_n}{\rightarrow} u_{n+1}$$

En general, y como no hay confusión, escribiremos $u\in \rho$ para expresar que $u\in sop_V(\rho)$, y $e\in \rho$ para expresar que $e\in sop_E(\rho)$. 

\begin{remark}
	\quad 
	\begin{itemize}
	\item Siguiendo una notación similar al caso de las aristas binarias dirigidas, si $\rho\in \mathcal{P}(G)$ y $\conecta{u}{\rho}{v}$, entonces escribiremos $\rho^o=u$ y $\rho^i=v$.
	\item Cuando sea necesario, notaremos los caminos que pasan por $u$, que comienzan en $u$, y que acaban en $u$, respectivamente, por:
	$$\mathcal{P}_u(G)=\{\rho \in \mathcal{P}(G):\ u\in \rho\}$$
	$$\mathcal{P}_u^o(G)=\{\rho \in \mathcal{P}(G):\ \rho^o=u\}$$
	$$\mathcal{P}_u^i(G)=\{\rho \in \mathcal{P}(G):\ \rho^i=u\}$$
\end{itemize}
\end{remark}
\section{Generalized Graph Query}

A continuación presentamos \emph{Generalized Graph Query} (GGQ, para abreviar, a partir de ahora), nuestra propuesta para llevar a cabo consultas de patrones en grafos. Teniendo en cuenta las diversas clasificaciones apuntadas anteriormente, podemos decir que esta propuesta permite llevar a cabo consultas estructurales y semánticas, exactas, óptimas, y basadas en un tipo de Regular Pattern Matching que permite, además de proyectar aristas del patrón en caminos (no necesariamente aristas) que cumplan las restricciones impuestas, expresar restricciones más complejas sobre cada elemento del patrón y realizar consultas que posean ciclos.

Una de las características que buscamos en nuestra herramienta es que permita obtener (de alguna manera) patrones complementarios a un patrón dado. Esto significa que si una estructura no verifica un patrón debe verificar siempre uno de sus patrones complementarios. Como hemos visto en la sección anterior, muchas de las herramientas desarrolladas para llevar a cabo consultas de patrones en grafos exigen que se cumpla una proyección entre el patrón y la estructura a evaluar. Dicha proyección impide evaluar la no existencia de elementos, algo que vamos a necesitar para generar estos patrones complementarios, por lo que nuestra propuesta no se basa en una proyección a la hora de verificar si una estructura cumple con un patrón determinado, sino que será construida en base a predicados lógicos, que facilitarán la generación de patrones complementarios.

Hemos de indicar que nuestro objetivo principal es el de proporcionar una formalización completa del modelo (frente a implementaciones incompletas desde el punto de vista formal, pero operativas), pero con el objetivo secundario de proporcionar una implementación que sea utilizable desde un punto de vista práctico\footnote{\url{https://github.com/palmagro/ggq}} (aunque más como una prueba de concepto que como una herramienta profesional en esta primera etapa).

En busca de nuestros objetivos, nos apoyaremos en el concepto de Grafo de Selección visto anteriormente, ampliándolo para añadirle Regular Pattern Matching y algunas características adicionales que nos permitirán obtener una mayor potencia expresiva en los patrones que se pueden construir.

Como principales características diferenciadoras respecto de los sistemas de consulta vistas en el apartado anterior, podemos indicar que:

\begin{itemize}
    \item Los GGQ pueden contener ciclos. Será un problema posterior considerar implementaciones de los GGQ que manipulen los ciclos adecuadamente, considerar restricciones adicionales para asegurar ciertos niveles de eficiencia en su ejecución real, o preocuparse en la etapa de diseño de la consulta de crear un patrón que sea eficiente en la implementación disponible.

    \item Los GGQ pueden evaluar subgrafos. Recordemos que en los Grafos de Selección clásicos sólo es posible evaluar un único nodo que representa a la tabla \textit{target}. En el caso de los GGQ, los \textit{elementos fijos} (elementos que deben pertenecer al subgrafo bajo evaluación)  serán representados a través de un predicado que obliga a que dichos elementos estén contenidos en el subgrafo a evaluar.

    \item Las aristas individuales del GGQ pueden ser proyectadas sobre caminos en el grafo en el que se comprueba el patrón. Para ello se hará uso de predicados de forma similar a como se hace en Regular Pattern Matching.

    \item Los predicados asociados a nodos o aristas en el GGQ pueden evaluar características estructurales y semánticas más allá de las propiedades almacenadas a través de la función $\mu$ (por ejemplo, a través de métricas sobre el grafo o sus elementos).

\end{itemize}

Aunque ya hemos mencionado que podemos disponer de un conjunto de predicados asociados a los elementos del patrón, vamos a formalizar brevemente qué entendemos concretamente por un predicado definido sobre un grafo.

Tal y como muestra su definición, asociado a un grafo con propiedades tenemos una función $\mu$ que representa un conjunto de funciones (en particular, pueden ser predicados) asociadas a nodos y aristas del grafo. Consideremos $\Theta$, una colección de símbolos de función, predicados y constantes, que contiene todas las funciones de $\mu$ junto con constantes asociadas a cada elemento del grafo y, posiblemente, algunos símbolos adicionales, tanto de funciones como de predicados y constantes (por ejemplo, métricas definidas sobre los elementos del grafo). A partir de este conjunto de símbolos podemos definir un Lenguaje de Primer Orden con igualdad, $L$, haciendo uso de $\Theta$ como conjunto de símbolos no lógicos, sobre el que construimos, de la forma usual, el conjunto de términos del lenguaje y el conjunto de fórmulas, $FORM(L)$, que llamaremos \emph{predicados}.

Aunque, en general, las fórmulas definibles en $L$ se pueden aplicar a todos los objetos del universo, que en nuestro contexto estará compuesto por elementos de grafos (nodos, aristas, y estructuras formadas a partir de éstos), cuando queramos explicitar sobre qué tipos de objetos estamos trabajando en cada momento, podremos escribir $FORM_V(L)$ para indicar que son fórmulas aplicables sobre nodos, $FORM_E(L)$ para indicar que son fórmulas aplicables sobre aristas, $FORM_\mathcal{P}(L)$ para indicar que son fórmulas aplicables sobre caminos, etc.

En lo que sigue supondremos prefijado un Lenguaje sobre grafos, $L$, por lo que, con el objetivo de simplificar las expresiones que usemos, notaremos de forma general $FORM$ para denotar $FORM(L)$ cuando no haya posibilidad de confusión.

Además, y aprovechando la capacidad expresiva de los grafos generalizados, definimos las consultas sobre ellos haciendo uso de las mismas estructuras:

\begin{definition}
    Un \emph{Generalized Graph Query (GGQ)} sobre $L$ es un grafo binario con propiedades sobre $L$, $Q = (V_Q,E_Q,\mu_Q)$, donde existen $\alpha$ y $\theta$, propiedades destacadas en $\mu_Q$, tales que:
    \begin{itemize}
        \item $\alpha:V_Q\cup E_Q\rightarrow \{+,-\}$ total.
        \item $\theta:V_Q\cup E_Q\rightarrow FORM(L)$ asocia un predicado binario, $\theta_x$, a cada elemento $x$ de $V_Q\cup E_Q$.
    \end{itemize}
\end{definition}

Escribiremos $Q\in GGQ(L)$ para denotar que $Q$ es un Generalized Graph Query sobre $L$ (si el lenguaje está prefijado y no hay posibilidad de confusión, escribiremos simplemente $Q\in GGQ$).

El sentido de usar predicados binarios es que en la semántica asociada a un GGQ usaremos la segunda entrada de estos predicados para poder hablar de condiciones de pertenencia sobre subgrafos de $G$ (el grafo general sobre el que estamos evaluando las consultas), mientras que la primera esperará recibir como entrada elementos adecuados al tipo de elemento al que está asociado. Así, si $S$ es un subgrafo y $a\in V_Q$ entonces $\theta_a(.,S)\in FORM_V$, y si $e\in E_Q$ entonces $\theta_e(.,S)\in FORM_\mathcal{P}$. Por ejemplo:
\begin{align*}
\theta_a(v,S)&=\exists z\in S\ (\conecta{z}{}{v})\\
\theta_e(\rho,S)&=\exists y,z (\conecta{y}{\rho}{z} \wedge y \notin S\wedge z\in S)
\end{align*}
El primer predicado tendrá sentido para nodos, y se verificará cuando exista un camino en $G$ que conecta un nodo de $S$ (el subgrafo que estamos evaluando) con $v$, el nodo de entrada sobre el que se evalúa. El segundo predicado tendrá sentido para caminos, y se verificará cuando el camino evaluado, $\rho$, conecta $S$ con su complementario (en $G$).

Dado un GGQ en las condiciones anteriores, notaremos $x^+$, respectivamente $x^-$, para indicar que $\alpha(x)=+$, respectivamente $\alpha(x)=-$, y $V_Q^+/V_Q^-$ (respectivamente, $E_Q^+/E_Q^-$ ) el conjunto de nodos (respectivamente, aristas) positivos/negativos. Si para un elemento $x$, $\theta_x$ no está explícitamente definida, supondremos que $\theta_x$ es una tautología, que podemos denotar en general por $T$.

Tal y como veremos a continuación, intuitivamente los elementos positivos del patrón representan elementos que deben estar presentes en el grafo sobre el que se realiza la consulta y que verifican los predicados asociados, mientras que los elementos negativos en el patrón representan elementos que no deben estar presentes en el grafo.

Para poder expresar con más facilidad las condiciones necesarias que definen la aplicación de un GGQ sobre un grafo, así como los resultados que veremos más adelante, introducimos a continuación una serie de notaciones que generan predicados aplicables sobre elementos del grafo:

\begin{definition}
    Dado $Q=(V_Q,E_Q,\mu_Q)$ un GGQ, el conjunto de $Q$-predicados asociados a $Q$ es:
    \begin{enumerate}
        \item Para cada arista, $e\in E_Q$, definimos los $Q$-predicados asociados como:
        $$Q_{e^o}(v,S)=\exists \rho\in \mathcal{P}_v^o(G)\ \left(\theta_e(\rho,S)\wedge \theta_{e^o}(\rho^o,S) \wedge \theta_{e^i}(\rho^i,S)\right)$$
        $$Q_{e^i}(v,S)=\exists \rho\in \mathcal{P}_v^i(G)\ \left(\theta_e(\rho,S)\wedge \theta_{e^o}(\rho^o,S) \wedge \theta_{e^i}(\rho^i,S)\right)$$

        En general, escribiremos $Q_{e^*}(v,S)$, donde $*\in \{o,i\}$, y notaremos:
        $$Q_{e^*}^+ = Q_{e^*},\hspace{1cm} Q_{e^*}^- = \neg Q_{e^*}$$
        \item Para cada nodo, $n\in V_Q$, definimos el $Q$-predicado asociado como:
        \begin{align*}
        Q_n(S)&=\exists v\in V\ \left(\bigwedge_{e\in \gamma^o(n)} Q_{e^o}^{\alpha(e)}(v,S)\ \wedge \bigwedge_{e\in \gamma^i(n)} Q_{e^i}^{\alpha(e)}(v,S)\right)\\
        & = \exists v\in V\ \left(\bigwedge_{e\in \gamma^*(n)} Q_{e^*}^{\alpha(e)}(v,S)\right)
        \end{align*}
        Y que podemos escribir en general como:
        $$Q_n(S)=\exists v\in V\ \left(\bigwedge_{e\in \gamma(n)} Q_{e}^{\alpha(e)}(v,S)\right)$$

        ya que para cada nodo no hay posibilidad de confusión.
        Además, notaremos:
        $$Q_n^+ = Q_n,\hspace{1cm} Q_n^- = \neg Q_n$$
    \end{enumerate}
\end{definition}

A partir de estas notaciones, podemos definir formalmente cuándo un subgrafo verifica un GGQ determinado:

\begin{definition}
\label{verifica}
    Dado un subgrafo $S$ de un grafo con propiedades, $G=(V,E,\mu)$, y un Generalized Graph Query, $Q=(V_Q,E_Q,\mu_Q)$, ambos sobre el lenguaje $L$, diremos que $S$ \emph{verifica} $Q$, y lo denotaremos $S\vDash Q$, si se verifica la fórmula:

    $$Q(S)=\bigwedge_{n\in V_Q} Q_n^{\alpha(n)}(S)$$

    En caso contrario, escribiremos: $S\nvDash Q$.
\end{definition}

\begin{figure}[h!]
    \begin{center}
        \includegraphics[scale=0.4]{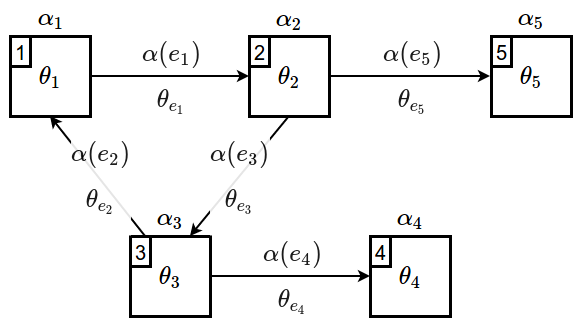}
    \end{center}
    \caption{%
        Ejemplo de Generalized Graph Query.
    }%
    \label{abstractpqg}
\end{figure}

En la Figura \ref{abstractpqg} se muestra un GGQ genérico a modo de ejemplo.

Uno de los objetivos que persiguen los GGQ es proporcionar la capacidad expresiva suficiente para expresar condiciones que hacen uso de elementos que están fuera del subgrafo que se está evaluando, algo que se ha demostrado necesario para disponer de un lenguaje de consultas potente y que, salvo en los grafos de selección, y de forma muy limitada, no está presente en el resto de soluciones vistas anteriormente.

Obsérvese que, en particular, usando $S=G$ podemos definir cuándo un grafo verifica un GGQ.

Aunque la definición de GGQ que hemos presentado hace uso de grafos binarios (no hipergrafos), ya que proyecta aristas sobre caminos que conectan pares de nodos, el concepto de grafo generalizado es suficientemente flexible como para permitir otras interpretaciones en las que se pueden considerar GGQs que hagan uso de estructuras más generales. Además, y es importante resaltar este hecho, aunque un GGQ sea binario, puede aplicarse sobre grafos con propiedades que no lo sean (es decir, $G$ podría ser un hipergrafo generalizado), ya que el concepto de camino que conecta pares de nodos se define independientemente de la aridad de las aristas que intervienen. En estos casos, se debería usar una notación algo más compleja para poder definir los $Q$-predicados, pero es completamente factible. Por motivos de simplicidad, y por la falta de bases de datos de hipergrafos, hemos restringido las definiciones presentadas a estos casos particulares, pero quedan abiertas para ser extendidas a los casos más generales en el momento en el que el uso de hipergrafos se generalice como medio de modelado y almacenamiento, ya que en la mayoría de las (escasas) ocasiones en que se han necesitado siempre se ha resuelto el problema por medio de la creación de nuevos tipos de nodos y aristas binarias que simulan la presencia de hiperaristas.

Antes de pasar a analizar algunas propiedades interesantes sobre los GGQ y la forma de construirlos, veamos algunos ejemplos que permitan entender cómo se interpretan y qué capacidad expresiva permiten.
\section{Ejemplos Representativos}
A lo largo de este parágrafo, y a modo de ejemplo, presentaremos una colección de pequeños GGQ sobre un grafo con propiedades concreto con el objeto de mostrar la forma en que funcionan y su capacidad expresiva.

\begin{figure}[h!]
    \begin{center}
        \includegraphics[width=\textwidth]{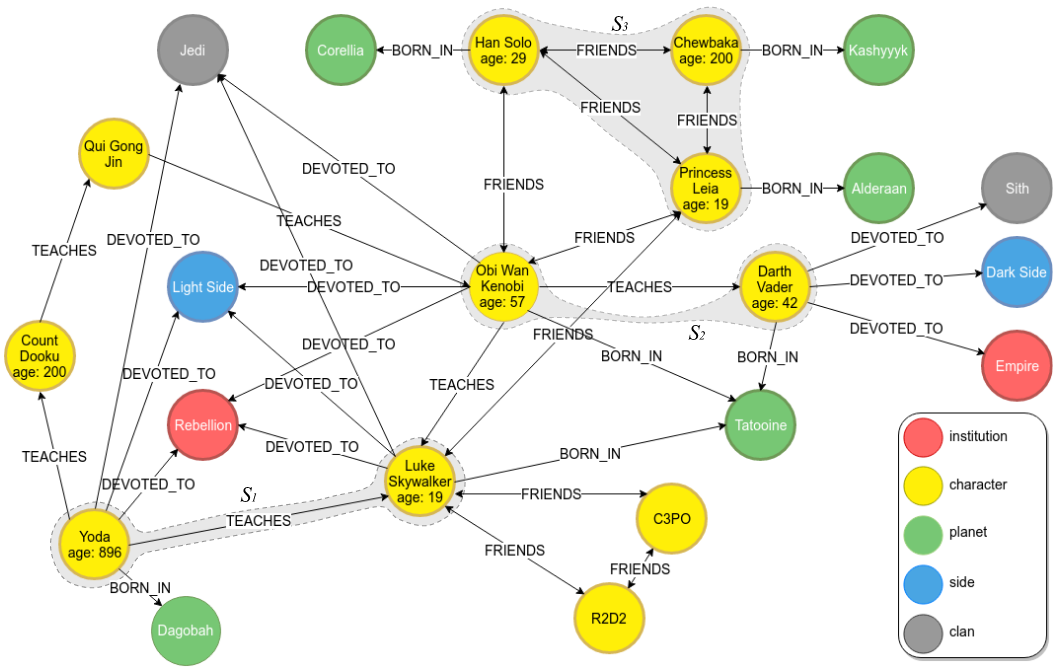}
    \end{center}
    \caption{%
        Grafo Starwars para ilustrar ejemplos de Generalized Graph Query.
    }%
    \label{starwars}
\end{figure}

En la Figura \ref{starwars} se presenta un grafo con propiedades que se corresponde con una sección de una base de datos basada en grafos que contiene información acerca de los personajes principales de la serie Starwars y que es utilizada frecuentemente como ejemplo sencillo para hacer demostraciones relacionadas con las capacidades de las bases de datos en grafo \footnote{http://console.neo4j.org/?id=StarWars}. En lo que sigue haremos uso de este grafo para presentar algunos patrones que hagan uso del lenguaje sobre el que está definido y para comprobar la verificación de algunos subgrafos concretos del mismo.

Con el fin de simplificar la representación de consultas y subgrafos, una de las propiedades en $\mu$, a la que denominaremos $\tau$ y que representa una clasificación de tipos sobre nodos y aristas, será expresada directamente sobre la aristas y, en el caso de los nodos, a través de colores. Además, la propiedad \textit{name} de los nodos será representada directamente sobre los mismos, y las aristas no dirigidas serán representadas como aristas bidireccionales.

\begin{figure}[h!]
    \begin{center}
        \includegraphics[scale=0.35]{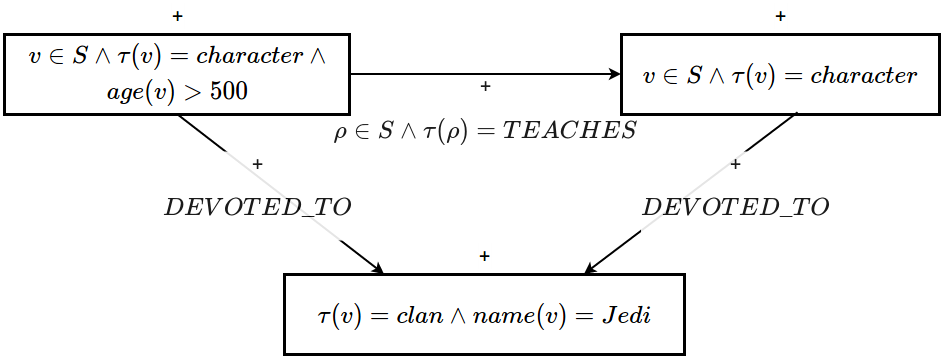}
    \end{center}
    \caption{%
        Ejemplo 1 GGQ.
    }%
    \label{pqg1}
\end{figure}

\begin{figure}[h]
    \begin{center}
        \includegraphics[scale=0.35]{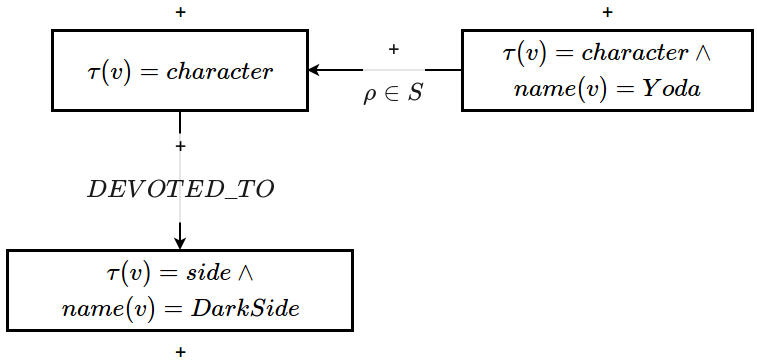}
    \end{center}
    \caption{%
        Ejemplo 6 GGQ.
    }%
    \label{pqg6}
\end{figure}

La representación gráfica de los GGQ de ejemplo se muestra en las figuras \ref{pqg1} a \ref{pqg6}. Cuando analicemos la interpretación de estas consultas también indicaremos algunos subgrafos de $G$ que los verifican. Cada elemento en estos GGQ tiene asociada la representación de su propiedad $\alpha$ directamente por medio de un símbolo $+/-$, y de su propiedad $\theta$ directamente en el elemento (si el predicado asociado a un elemento del GGQ es una tautología, dicho predicado no será representado). En expresiones del tipo $\tau(\rho) = X$ en el predicado de una arista, $X$ se  interpreta como una expresión regular que debe verificarse por la secuencia de propiedades $\tau$ de $sop_E(\rho)$.

El GGQ $P_1$ (Figura \ref{pqg1}) se puede interpretar en lenguaje natural a través de la siguiente sentencia: \textit{Personajes y relación alumno-maestro en la que ambos son devotos de los Jedi y el maestro tiene más de 500 años}. En este caso se imponen restricciones estructurales a través de la presencia de aristas y a través de predicados que hacen uso de las propiedades $\tau$, \textit{name}, y \textit{age}. Este GGQ se verificará en subgrafos en los que puedan ser proyectados dos nodos y una arista que los une (los tres elementos marcados como elementos positivos en el GGQ) que cumplan con las restricciones impuestas. En el caso de que existiera un personaje que se haya enseñado a sí mismo (lo que vendría dado por un lazo de tipo \texttt{TEACHES}) que tenga más de 500 años y sea devoto de los Jedi, un subgrafo que contenga este nodo también verificaría este patrón. El subgrafo marcado como $S_1$ en la Figura  \ref{starwars} verifica $P_1$.

\begin{figure}[h!]
    \begin{center}
        \includegraphics[scale=0.35]{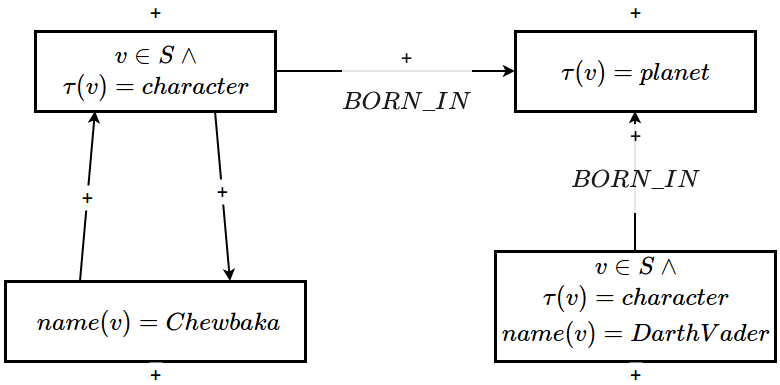}
    \end{center}
    \caption{%
        Ejemplo 2 GGQ.
    }%
    \label{pqg2}
\end{figure}

El GGQ $P_2$ (Figura \ref{pqg2}) se puede interpretar en lenguaje natural a través de la siguiente sentencia: \textit{subgrafos que contengan a Darth Vader y a un personaje que provenga del mismo planeta que él y que posea un camino que lo conecte con Chewbaka}. Este GGQ presenta un nodo positivo que representa al personaje \textit{Chewbaka} (por medio de la propiedad $name$) e impone una restricción que exige que uno de los nodos en el subgrafo evaluado esté conectado con él. Un subgrafo que verifica $P_2$ es el subgrafo destacado como $S_2$ en la Figura \ref{starwars}.

\begin{figure}[h]
    \begin{center}
        \includegraphics[scale=0.35]{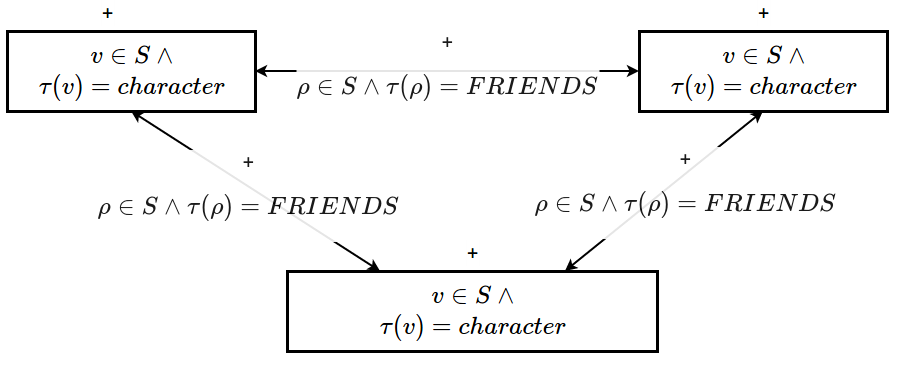}
    \end{center}
    \caption{%
       Ejemplo 3 GGQ.
    }%
    \label{pqg3}
\end{figure}

El GGQ $P_3$ (Figura \ref{pqg3}) presenta un ciclo a través de relaciones de amistad, y $S_3$ (resaltado en la Figura \ref{starwars}) es un subgrafo que lo verifica. Cualquier subgrafo que contenga tres personajes que son amigos entre sí (existen relaciones de tipo \texttt{FRIENDS} entre ellos) verificará $P_3$. Por ejemplo, subgrafos que contengan el ciclo formado por \texttt{Luke Skywalker}, \texttt{R2D2} y \texttt{C3PO}, o el ciclo formado por \texttt{Han Solo}, \texttt{Princess Leia} y \texttt{Chewbaka}.

\begin{figure}[h]
    \begin{center}
        \includegraphics[scale=0.35]{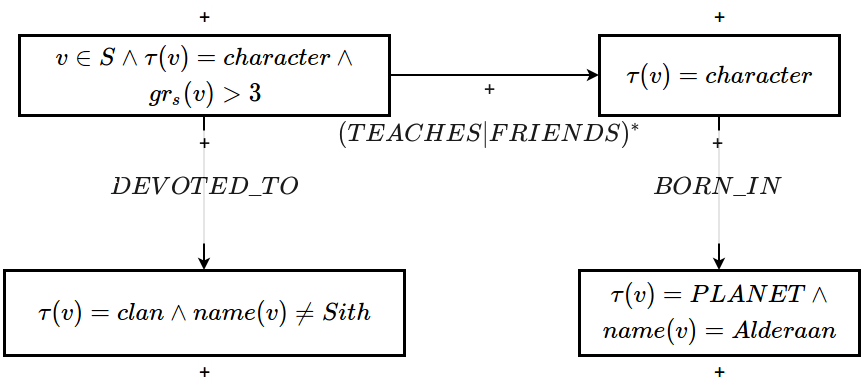}
    \end{center}
    \caption{%
        Ejemplo 4 GGQ.
    }%
    \label{pqg4}
\end{figure}

El GGQ $P_4$ (Figura \ref{pqg4}) puede ser interpretado en lenguaje natural a través de la siguiente sentencia: \textit{Personaje que esté conectado a través de relaciones de tipo \texttt{FRIENDS} o \texttt{TEACHES} con alguien que provenga de Alderaan, que tenga grado de salida superior a tres, y que sea devoto de un clan que no sean los Sith}. En este caso, se ha utilizado una expresión regular para expresar un camino compuesto de relaciones de tipo \texttt{FRIENDS} o \texttt{TEACHES}, además se ha usado una función auxiliar, $ gr_s (v) $, para referirse al grado de salida del nodo $v$. Cualquier subgrafo conteniendo el nodo \texttt{Luke Skywalker} o el nodo \texttt{Obi Wan Kenobi} verificará $P_4$.

\begin{figure}[h]
    \begin{center}
        \includegraphics[scale=0.35]{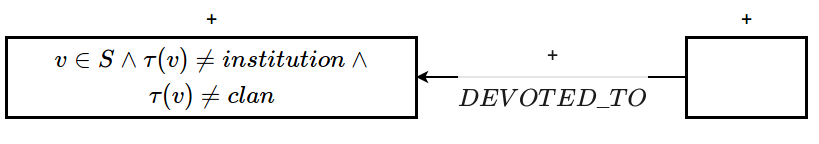}
    \end{center}
    \caption{%
       Ejemplo 5 GGQ.
    }%
    \label{pqg5}
\end{figure}

El GGQ $P_5$ (Figura \ref{pqg5}) representa la consulta: \textit{Nodos que no sean instituciones ni clanes, pero tengan devotos}, y sólo es verificado por los nodos de tipo \textit{side}. En este caso, se ha utilizado un nodo cuya propiedad $\theta$ es una tautología (que, como indicamos, ha sido representado a través de un nodo vacío).

El GGQ $P_6$ (Figura \ref{pqg6}) podría ser interpretado a través de la siguiente sentencia: \textit{Caminos que relacionen a Yoda con personajes del Lado Oscuro}. Cualquier subgrafo que contenga el camino (\texttt{Yoda}) $\rightarrow$ (\texttt{Count Dooku})  $\rightarrow$ (\texttt{Qui Gong Jin}) $\rightarrow$ (\texttt{Obi Wan Kenobi}) $\rightarrow$ (\texttt{Darth Vader}) verificará $ P_6 $.
\section{Conjuntos de Refinamiento}

En las secciones anteriores hemos visto que los GGQ se pueden interpretar como predicados sobre la familia de subgrafos de un grafo prefijado $G$. Sería interesante obtener formas computacionalmente efectivas de construir GGQ a partir de operaciones básicas para obtener familias de predicados que nos permitan analizar la estructura de los subgrafos de $G$ de una forma automática.

A continuación vamos a dar una primera aproximación a un método constructivo que sea de utilidad para realizar este tipo de tareas sobre un grafo. Comenzamos dando una definición natural cuando se trabaja con consultas sobre estructuras, y que nos permite ver cuándo un GGQ tiene más capacidad que otro para discriminar entre subgrafos.

\begin{definition}
    Dados $Q_1,\ Q_2\in GGQ$, diremos que $Q_1$ \emph{refina} $Q_2$ en $G$, y lo notaremos como $Q_1\preceq_G Q_2$ (escribiremos, simplemente, $\preceq$ cuando trabajemos sobre un grafo $G$ prefijado) si:

    $$\forall S\subseteq G\ (S\vDash Q_1 \Rightarrow S\vDash Q_2)$$
\end{definition}

\begin{figure}[h]
    \begin{center}
        \includegraphics[scale=0.33]{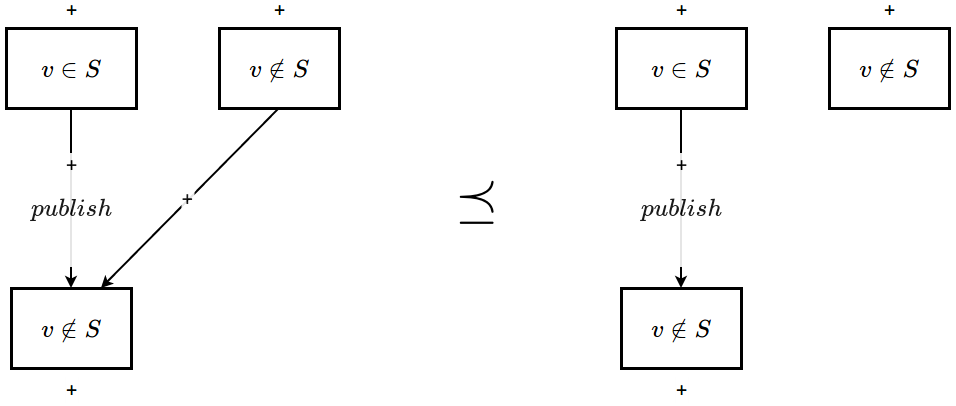}
    \end{center}
    \caption{%
        Refinado entre GGQ.
    }%
    \label{ejemplo-refinar}
\end{figure}

En la Figura \ref{ejemplo-refinar} se muestra un ejemplo de un GGQ que refina a otro. En este caso, el GGQ de la izquierda refina al GGQ situado a la derecha ya que, además de exigir que uno de los nodos en el subgrafo a evaluar esté conectado con un nodo que no pertenece a dicho subgrafo a través de una relación de tipo \texttt{publish}, también se exige que el nodo destino de dicha relación posea una arista entrante que parta de un nodo que no pertenece al subgrafo bajo evaluación.

De forma natural, podemos definir cuándo dos GGQ son equivalentes como consultas, que se tendrá cuando los dos verifiquen exactamente los mismos subgrafos.

\begin{definition}
    Dados $Q_1,\ Q_2\in GGQ$, diremos que son \emph{equivalentes} en $G$, y lo notaremos como $Q_1\equiv_G Q_2$ (usaremos, simplemente, $\equiv$ cuando trabajemos sobre un grafo $G$ prefijado) si:

    $$Q_1\preceq_G Q_2\ \wedge \ Q_2\preceq_G Q_1$$
\end{definition}

\begin{figure}[h]
    \begin{center}
        \includegraphics[scale=0.33]{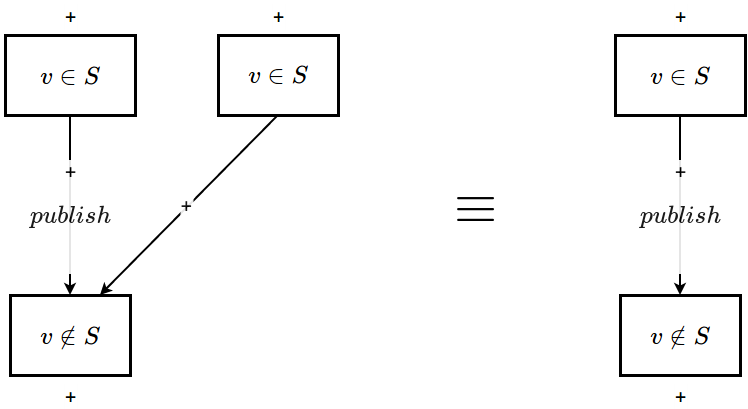}
    \end{center}
    \caption{%
        Equivalencia de GGQ.
    }%
    \label{ejemplo-equivalencia}
\end{figure}

En la Figura \ref{ejemplo-equivalencia} se muestra un ejemplo de dos GGQ que son equivalentes. Nótese que la única diferencia estructural del GGQ de la izquierda con respecto al de la derecha es la existencia de un nodo perteneciente al subgrafo a evaluar que posea una arista de salida cuyo destino sea un nodo que no esté incluido en el subgrafo a evaluar. Dicha restricción está incluida en el GGQ de la derecha, ya que de existir un nodo en el subgrafo bajo evaluación con una arista de tipo \texttt{publish} conectada con un nodo que no pertenezca a dicho subgrafo, también existe una arista con las mismas restricciones excepto el tipo impuesto a la arista.

Es fácil probar el siguiente resultado, que nos dice que $\preceq_G$ genera una relación de orden sobre los GGQ considerando la equivalencia como igualdad (o lo que es lo mismo, trabajando en el espacio cociente que determina la equivalencia).

\begin{theorem}
    Para todo Grafo con Propiedades, $G$, se tiene que $(GGQ, \preceq_G)$ es un conjunto ordenado. Es decir:
    \begin{enumerate}
        \item $\forall Q\in GGQ\ (Q\preceq_G Q)$.
        \item $Q_1\preceq_G Q_2\ \wedge \ Q_2\preceq_G Q_1 \Rightarrow Q1\equiv_G Q2$.
        \item $Q_1\preceq_G Q_2\ \wedge \ Q_2\preceq_G Q_3 \Rightarrow Q_1\preceq_G Q_3$.
    \end{enumerate}
\end{theorem}

Es fácil comprobar que, en general, $\preceq_G$ no genera una relación de orden total. Para ello, basta encontrar dos GGQ, $Q_1$ y $Q_2$, para los cuales no se cumpla $(Q_1 \rightarrow Q_2) \lor  (Q_2 \rightarrow Q_1)$. Por ejemplo, si $Q_1$ es un GGQ compuesto por un único nodo positivo con la restricción $v \in S$, y $Q_2$ es un GGQ compuesto también por un único nodo positivo con la restricción $v \notin S$, entonces $Q_1$ exigirá que el subgrafo bajo evaluación no esté vacío, y $Q_2$ exigirá que exista algún nodo en el grafo que no pertenezca al subgrafo bajo evaluación. Ambas restricciones son independientes por lo que no existe implicación entre los predicados que los GGQ representan.

Vamos a analizar la relación existente entre la estructura topológica de un GGQ y su funcionalidad como predicado sobre subgrafos. En general, es complicado intentar extraer propiedades lógicas del predicado a partir de las propiedades estructurales del grafo que lo representa, pero podemos obtener algunas condiciones útiles que nos permitirán manipular constructivamente las estructuras para modificar la interpretación de los GGQ de forma controlada.

\begin{definition}
    Dados $Q_1,\ Q_2\in GGQ$, diremos que $Q_1$ es una \emph{extensión $Q^-$-conservativa} de $Q_2$, y lo notaremos como $Q_2\subseteq^- Q_1$, si:
    \begin{enumerate}
        \item $Q_2\subseteq Q_1$ (como grafos con propiedades, por lo que en los elementos de $Q_2$ coinciden los valores de $\alpha$ y $\theta$ para ambos GGQ).
        \item Para cada nodo negativo de $Q_2$, $n\in V_{Q_2}^-$, y cada arista incidente en él en $Q_1$, $e \in \gamma_{Q_1}(n)$, existe una arista incidente en él en $Q_2$, $e'\in \gamma_{Q_2}(n)$, que impone la misma restricción, es decir: $Q_e\equiv Q_{e'}$.
    \end{enumerate}
\end{definition}

\begin{figure}[h]
    \begin{center}
        \includegraphics[scale=0.33]{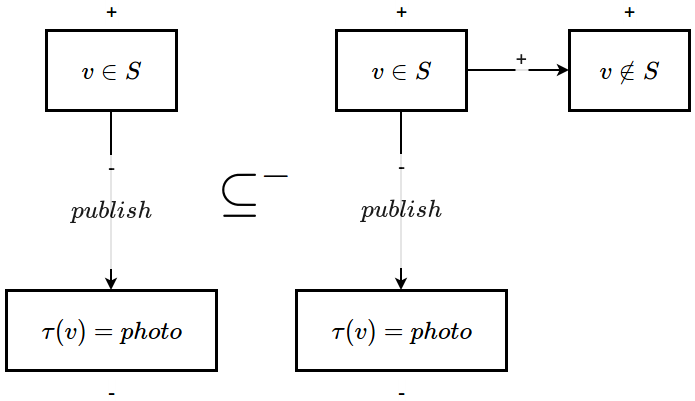}
    \end{center}
    \caption{%
        Extensión $Q^-$-conservativa.
    }%
    \label{ejemplo-conservativa}
\end{figure}

La Figura \ref{ejemplo-conservativa} muestra un ejemplo de extensión $Q^-$-conservativa. La extensión realizada en el GGQ de la izquierda para obtener el GGQ de la derecha impone nuevas restricciones sobre el nodo positivo pero no añade nuevas restricciones al nodo negativo. 

Como los nodos negativos añaden restricciones de no existencia a la verificación de subgrafos, las extensiones $Q^-$-conservativas aseguran que no estamos añadiendo restricciones adicionales a éstos (añadiendo más información a sus aristas incidentes), por lo que podemos dar el siguiente resultado:

\begin{theorem}
    Dados $Q_1,Q_2\in GGQ$, si $Q_2\subseteq^- Q_1$ entonces $Q_1\preceq Q_2$.
\end{theorem}

\begin{proof}
    Debido a que los $Q$-predicados para aristas dependen exclusivamente de la información en la propia arista (que considera el valor de $\theta$ en sus extremos, sin importar el valor de $\alpha$ en los mismos), podemos afirmar que:
    $$\forall e\in E_{Q_2}\ ({Q_1}_e^{\alpha(e)}={Q_2}_e^{\alpha(e)})$$

    Teniendo en cuenta este hecho, vamos a analizar cómo se comportan los $Q$-predicados asociados a los nodos para ambos GGQ:
    \begin{itemize}
        \item Si  $n\in V_{Q_2}^-$, debido a que $Q_2\subseteq^- Q_1$, es inmediato que ${Q_1}_n^-={Q_2}_n^-$ (es la misma fórmula, no solo son equivalentes).
        \item Si $n\in V_{Q_2}^+$, entonces ${Q_1}_n^+\rightarrow {Q_2}_n^+$, ya que (notaremos por $\gamma_1$, $\gamma_2$ las funciones de incidencia de $Q_1$ y $Q_2$, respectivamente):
        \begin{align*}
        {Q_1}_n^+   &= \exists v\in V\ \left( \bigwedge_{e\in \gamma_1(n)} {Q_1}_e^{\alpha(e)}\right)\\
                &= \exists v\in V\ \left( \bigwedge_{e\in \gamma_1(n)\cap E_{Q_2}} {Q_1}_e^{\alpha(e)}\ \wedge \ \bigwedge_{e\in \gamma_1(n)\smallsetminus E_{Q_2}} {Q_1}_e^{\alpha(e)}\right)\\
                &= \exists v\in V\ \left( \bigwedge_{e\in \gamma_2(n)\cap E_{Q_2}} {Q_2}_e^{\alpha(e)}\ \wedge \ \bigwedge_{e\in \gamma_1(n)\smallsetminus E_{Q_2}} {Q_1}_e^{\alpha(e)}\right)\\
                &\rightarrow {Q_2}_n^+
        \end{align*}
    \end{itemize}
    En consecuencia:
    \begin{align*}
    Q_1 &= \bigwedge_{n\in V_{Q_1}} {Q_1}_n^{\alpha(n)} = \bigwedge_{n\in V_{Q_2}} {Q_1}_n^{\alpha(n)} \ \wedge \ \bigwedge_{n\in V_{Q_1}\smallsetminus V_{Q_2}} {Q_1}_n^{\alpha(n)}\\
    &= \bigwedge_{n\in V_{Q_2}^+} {Q_1}_n^{\alpha(n)} \ \wedge \ \bigwedge_{n\in V_{Q_2}^-} {Q_1}_n^{\alpha(n)} \ \wedge \ \bigwedge_{n\in V_{Q_1}\smallsetminus V_{Q_2}} {Q_1}_n^{\alpha(n)}
\end{align*}
    \begin{align*}
    &\rightarrow \bigwedge_{n\in V_{Q_2}^+} {Q_2}_n^{\alpha(n)} \ \wedge \ \bigwedge_{n\in V_{Q_2}^-} {Q_2}_n^{\alpha(n)} \ \wedge \ \bigwedge_{n\in V_{Q_1}\smallsetminus V_{Q_2}} {Q_1}_n^{\alpha(n)}\\ 
    &= \bigwedge_{n\in V_{Q_2}} {Q_2}_n^{\alpha(n)} \ \wedge \ \bigwedge_{n\in V_{Q_1}\smallsetminus V_{Q_2}} {Q_1}_n^{\alpha(n)}\\
    &\rightarrow {Q_2}
    \end{align*}
\end{proof}

El resultado anterior sugiere que se puede refinar un GGQ añadiendo nodos (de cualquier signo) y aristas a los nodos positivos ya existentes, pero debido a la interpretación (negada) de los $Q$-predicados asociados a nodos negativos, hay que tener la precaución de mantener el entorno de los mismos para estar seguros de que añadir más aristas a ellos no debilita las condiciones impuestas a los subgrafos evaluados (y, por tanto, no conseguiríamos predicados que refinan).

Con el fin de obtener métodos controlados de generación de consultas, en lo que sigue daremos un método constructivo para ir refinando un GGQ por pasos unitarios. Para ello, comenzaremos viendo cómo se comportan los GGQ cuando se clonan nodos.

Un clon consiste en hacer copias de nodos existentes, clonando todas las aristas incidentes en ellos (y entre ellos, en caso de que clonemos varios nodos que están conectados en el GGQ original). Por supuesto, la operación de clonación se puede hacer sobre grafos con propiedades cualesquiera, y así la presentamos.

\begin{definition}
    Dado $G=(V,E,\mu)$ un grafo con propiedades, y $W\subseteq V$, definimos el \emph{clon de $G$ por duplicación de $W$}, y lo notaremos por $Cl_G^W$, como el grafo con propiedades siguiente:

    $$Cl_G^W=(V\cup W',\ E\cup E',\mu\cup \{(n',\mu(n))\}_{n\in W}\cup \{(e',\mu(e))\}_{e'\in E'})$$
    donde:
    \begin{itemize}
        \item para cada $n\in W$, $n'$ es un nodo nuevo, $W'=\{n'\ :\ n\in W\}$, y
        \item $E'$ es un conjunto de aristas nuevas que se consiguen a partir de las aristas incidentes en nodos de $W$ donde se sustituyen de todas las formas posibles los nodos de $W$ por copias de $W'$ (de forma que aparecen aristas clonadas que conectan nodos originales con nodos copia, y también aristas clonadas que conectan nodos copia).
    \end{itemize}
\end{definition}

\begin{figure}[h]
    \begin{center}
        \includegraphics[scale=0.33]{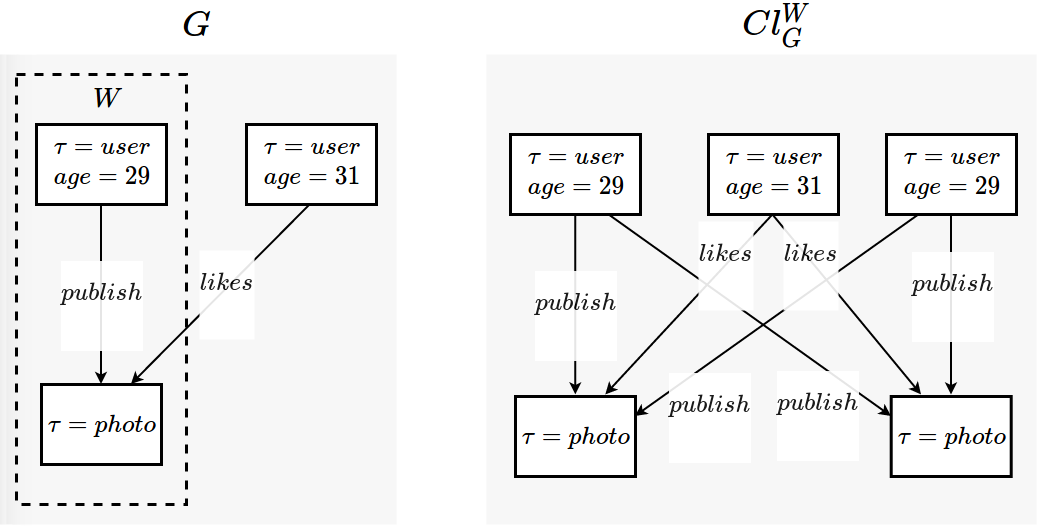}
    \end{center}
    \caption{%
        Clon de un grafo.
    }%
    \label{ejemplo-clonacion}
\end{figure}

La Figura \ref{ejemplo-clonacion} muestra un ejemplo de un grafo clonado por dupliación de dos de sus nodos. En el grafo original, a la izquierda, se resaltan los dos nodos a ser clonados. El resultado de la clonación se presenta en el grafo de la derecha. 

El siguiente resultado nos indica que la clonación de nodos positivos no altera la interpretación de las consultas.

\begin{theorem}
    Si $Q\in GGQ$ y $W\subseteq V_Q^+$, entonces $Cl_Q^W\equiv Q$.
\end{theorem}
\begin{proof}
    Para facilitar la notación, sea $Q_1=Cl_Q^W$. Entonces, siguiendo un razonamiento similar al de la demostración anterior:
    \begin{align*}
    Q_1 &= \bigwedge_{n\in V_{Q_1}} {Q_1}_n^{\alpha(n)}\\
    & =\bigwedge_{n\in V_Q} {Q_1}_n^{\alpha(n)}\ \wedge \ \bigwedge_{n\in W} {Q_1}_{n'}^{\alpha(n')}\\
    &= \bigwedge_{n\in V_Q\smallsetminus \gamma_Q(W)} {Q_1}_n^{\alpha(n)}\ \wedge \ \bigwedge_{n\in\gamma_Q(W)} {Q_1}_n^{\alpha(n)}\ \wedge \ \bigwedge_{n\in W} {Q_1}_{n'}^{\alpha(n')}\\
    &= \bigwedge_{n\in V_Q\smallsetminus \gamma_Q(W)} Q_n^{\alpha(n)}\ \wedge \bigwedge_{n\in\gamma_Q(W)} Q_n^{\alpha(n)}\ \wedge \ \bigwedge_{n\in W} Q_n^{\alpha(n)}\\
    &= Q
    \end{align*}
\end{proof}

Siguiendo con la idea de obtener herramientas que nos permitan construir GGQ de manera automática, el concepto de \emph{refinamiento} que introducimos a continuación completa las operaciones que podemos hacer para refinar un GGQ. En cierta forma, un conjunto de refinamiento forma una partición por refinamientos de un GGQ dado.

\begin{definition}
    Dado $Q\in GGQ$. Diremos que $R\subseteq GGQ$ es un \emph{conjunto de refinamiento} de $Q$ en $G$ si verifica:
    \begin{enumerate}
        \item $\forall\ Q'\in R\ (Q'\preceq_G Q)$
        \item $\forall\ S\subseteq G\ (S\vDash Q\Rightarrow \exists !\ Q'\in R\ (S\vDash Q'))$
    \end{enumerate}
\end{definition}

\begin{figure}[h]
    \begin{center}
        \includegraphics[scale=0.20]{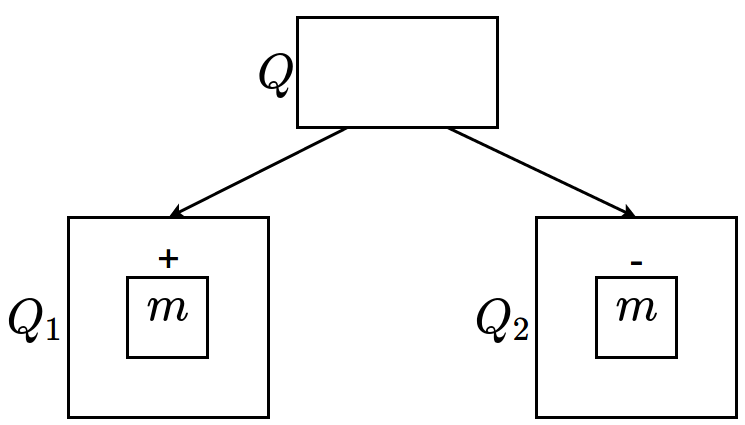}
    \end{center}
    \caption{%
        Refinamiento añadir nodo.
    }%
    \label{ref1}
\end{figure}

Estamos ya en condiciones de dar algunos conjuntos de refinamiento que nos permitirán automatizar los procesos de creación y modificación de Generalized Graph Queries. Comenzaremos por la operación más sencilla, que consiste en ver de qué formas se pueden añadir nuevos nodos a un GGQ existente:

\begin{theorem}[Añadir nodo nuevo a $Q$]
    Dado $Q\in GGQ$ y $m\notin V_Q$, entonces el conjunto que notaremos como $Q+\{m\}$, formado por:
        \begin{align*}
        Q_1 &= (V_Q\cup\{m\},\ E_Q,\ \alpha_Q\cup(m,+),\ \theta_Q\cup(m,T))\\
        Q_2 &= (V_Q\cup\{m\},\ E_Q,\ \alpha_Q\cup(m,-),\ \theta_Q\cup(m,T))
        \end{align*}
    es un conjunto de refinamiento de $Q$ en $G$ (Fig. \ref{ref1}).
\end{theorem}

\begin{proof}
    Hemos de comprobar que se verifican las dos condiciones necesarias para que sea un conjunto de refinamiento:
    \begin{enumerate}
        \item Es evidente que $Q\subseteq^- Q_1$ y $Q\subseteq^- Q_2$, por lo que $Q_1\preceq Q$ y $Q_2\preceq Q$.
        \item Sea $S\subseteq G$ tal que $S\vDash Q$. Tenemos que:
        \begin{align*}
        Q_1&= Q\ \wedge\ Q_m\\
        Q_2&= Q\ \wedge\ \neg Q_m
        \end{align*}
        donde $Q_m=\exists v\in V\ (T)$.

        Si $G\neq \emptyset$, entonces $S\vDash Q_1$ y $S\nvDash Q_2$.
        
        Si $G= \emptyset$, entonces $S\nvDash Q_1$ y $S\vDash Q_2$.
    \end{enumerate}
\end{proof}

Como norma general, $G\neq\emptyset$, por lo que esta operación realmente no refina, en el sentido de que $Q1\equiv Q$ y $Q2\equiv \neg T$. Sin embargo, a pesar de que obtenemos un GGQ equivalente, esta operación es muy útil para añadir nuevos nodos a un GGQ a los que posteriormente se le podrán ir añadiendo nuevas restricciones.

Teniendo en cuenta los resultados anteriores que daban relaciones entre las propiedades estructurales del GGQ y su interpretación semántica como consulta, pasamos a dar un segundo conjunto de refinamiento que nos indica cómo interviene la creación de aristas entre nodos existentes. Para mantener que todos refinen al GGQ original, hemos de restringir la adición de aristas a los nodos positivos.

\begin{figure}[h]
    \begin{center}
        \includegraphics[scale=0.30]{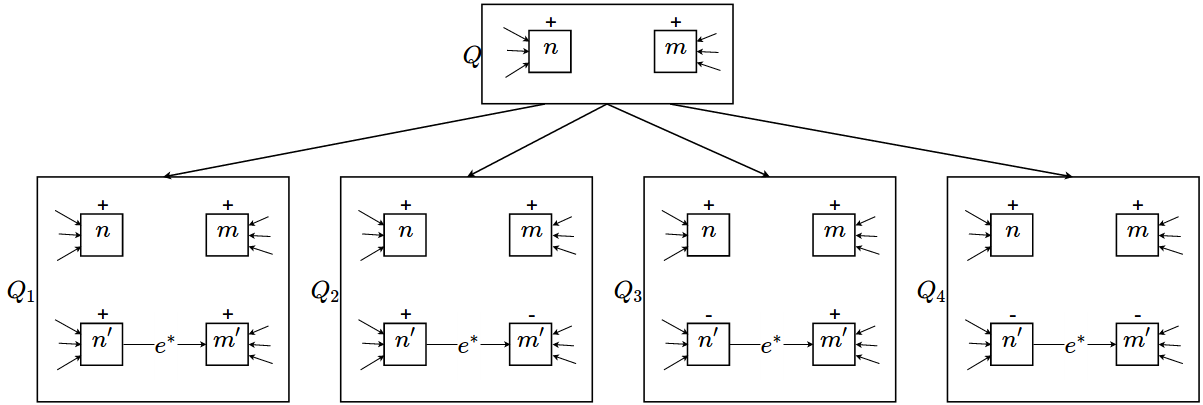}
    \end{center}
    \caption{%
        Refinamiento añadir arista.
    }%
    \label{ref2}
\end{figure}

\begin{theorem}[Añadir arista nueva entre nodos positivos de $Q$]

    Dado $Q\in GGQ$ y $n,m\in V_Q^+$, entonces el conjunto que denotaremos como $Q+\{n^+\overset {e^*}{\longrightarrow} m^+\}$ ($*\in\{+,-\}$), formado por (donde $Q'=Cl_Q^{\{n,m\}}$):
    \begin{align*}
    Q_1 &= (V_{Q'},\ E_{Q'}\cup\{n^+\overset {e^*}{\longrightarrow} m^+\},\ \theta_{Q'}\cup(e,T))\\
    Q_2 &= (V_{Q'},\ E_{Q'}\cup\{n^+\overset {e^*}{\longrightarrow} m^-\},\ \theta_{Q'}\cup(e,T))\\
    Q_3 &= (V_{Q'},\ E_{Q'}\cup\{n^-\overset {e^*}{\longrightarrow} m^+\},\ \theta_{Q'}\cup(e,T))\\
    Q_4 &= (V_{Q'},\ E_{Q'}\cup\{n^-\overset {e^*}{\longrightarrow} m^-\},\ \theta_{Q'}\cup(e,T))
\end{align*}
    es un conjunto de refinamiento de $Q$ en $G$ (Fig. \ref{ref2}).
\end{theorem}

\begin{proof}
    \quad
    \begin{enumerate}
        \item Como $Q'$ es un clon de $Q$, y $\{n,m\}\subseteq V_Q^+$, tenemos que $Q\equiv Q'$. Además, por construcción, $Q'\subseteq^- Q_1,Q_2,Q_3,Q_4$, por lo que $Q_1,Q_2,Q_3,Q_4\preceq Q'\equiv Q$.
        \item Consideremos los predicados:
        \begin{align*}
        P_n &= \exists v\in V\ \left( \bigwedge_{a\in \gamma(n)} Q_a^{\alpha(a)}\ \wedge Q_{e^o}^{\alpha(e)}\right)\\
        P_m &= \exists v\in V\ \left( \bigwedge_{a\in \gamma(m)} Q_a^{\alpha(a)}\ \wedge Q_{e^i}^{\alpha(e)}\right)
        \end{align*}
        Si $S\vDash Q_n$ y $S\vDash Q_m$, entonces tenemos 4 opciones mutuamente excluyentes, según se verifique $S\vDash P_n$ y/o $S\vDash P_m$, que son:
        \begin{itemize}
            \item $S\vDash P_n\ \wedge\ S\vDash P_m \Rightarrow S\vDash Q_1$
            \item $S\vDash P_n\ \wedge\ S\nvDash P_m \Rightarrow S\vDash Q_2$
            \item $S\nvDash P_n\ \wedge\ S\vDash P_m \Rightarrow S\vDash Q_3$
            \item $S\nvDash P_n\ \wedge\ S\nvDash P_m \Rightarrow S\vDash Q_4$
        \end{itemize}
    \end{enumerate}
\end{proof}
Si $n=m$ (la arista añadida es un lazo), entonces el conjunto de refinamiento anterior queda reducido a dos GGQ, los equivalentes a $Q_1$ y $Q_4$.

La siguiente modificación necesaria es la de añadir un predicado adicional a una arista existente. Para mantener las condiciones estructurales necesarias, restringimos esta operación a las aristas positivas que conectan nodos positivos.

\begin{figure}[h]
    \begin{center}
        \includegraphics[scale=0.30]{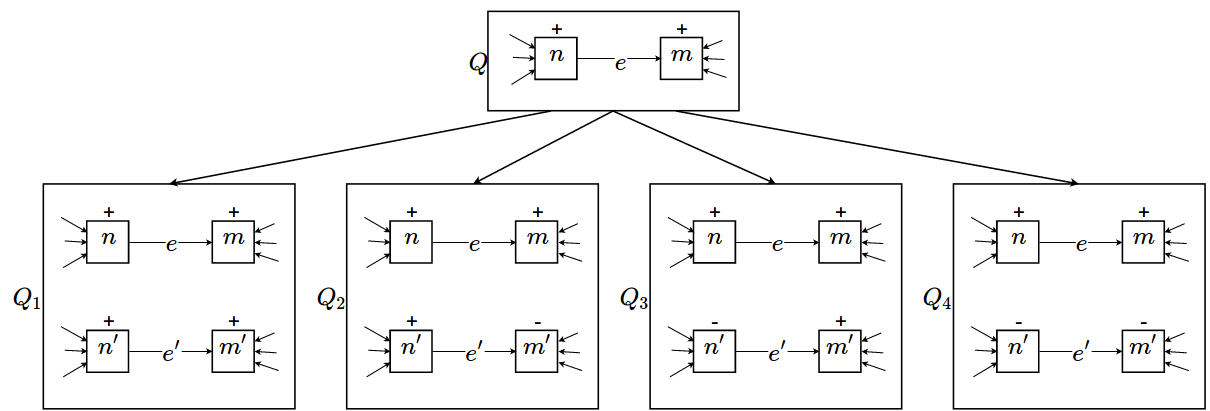}
    \end{center}
    \caption{%
        Refinamiento añadir predicado a arista.
    }%
    \label{ref3}
\end{figure}

\begin{theorem}[Añadir predicado a arista positiva entre nodos positivos de $Q$]

    Dado $Q\in GGQ$ $n,m\in V_Q^+$, con $n^+\overset {e^+}{\longrightarrow} m^+$, y $\varphi\in FORM$, el conjunto que notaremos como $Q+\{n^+\overset {e \wedge \varphi}{\longrightarrow} m^+\}$, formado por (donde $Q'=Cl_Q^{\{n,m\}}$):
    \begin{align*}
    Q1 &= (V_{Q'},\ E_{Q'}\cup\{n^+\overset {e'}{\longrightarrow} m^+\},\ \theta_{Q'}\cup(e',\theta_e\wedge \varphi))\\
    Q2 &= (V_{Q'},\ E_{Q'}\cup\{n^+\overset {e'}{\longrightarrow} m^-\},\ \theta_{Q'}\cup(e',\theta_e\wedge \varphi))\\
    Q3 &= (V_{Q'},\ E_{Q'}\cup\{n^-\overset {e'}{\longrightarrow} m^+\},\ \theta_{Q'}\cup(e',\theta_e\wedge \varphi))\\
    Q4 &= (V_{Q'},\ E_{Q'}\cup\{n^-\overset {e'}{\longrightarrow} m^-\},\ \theta_{Q'}\cup(e',\theta_e\wedge \varphi))
    \end{align*}
    es un conjunto de refinamiento de $Q$ en $G$ (Fig. \ref{ref3}).
\end{theorem}

\begin{proof}
    La demostración es similar a la realizada en los casos anteriores.
\end{proof}

Por último, la modificación que nos queda es la de añadir predicados a nodos existentes. De nuevo,
hemos de restringir esta operación a los casos que no plantean problemas, cuando los nodos afectados son positivos (el nodo al que se añade el predicado, y los conectados a él).

\begin{figure}[h]
    \begin{center}
        \includegraphics[scale=0.35]{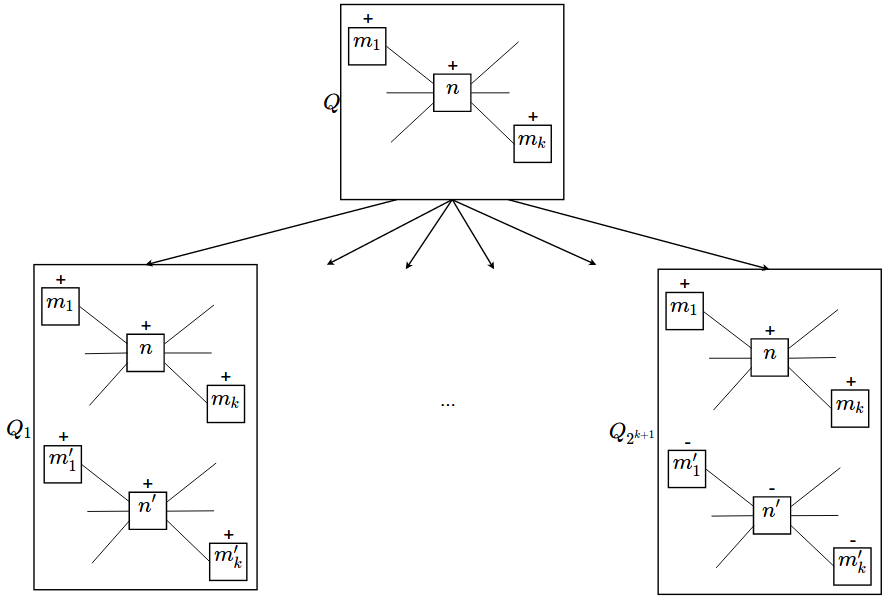}
    \end{center}
    \caption{%
        Refinamiento añadir predicado a nodo.
    }%
    \label{ref4}
\end{figure}

\begin{theorem}[Añadir predicado a nodo positivo con entorno positivo en $Q$]

    Dado $Q\in GGQ$, $n\in V_Q^+$, con $\mathcal{N}_Q(n)\subseteq V_Q^+$, y $\varphi\in FORM$. Definimos el conjunto que denotaremos como $Q+\{n\wedge \varphi\}$ formado por:
    $$\{Q_{\sigma}=(V_{Q'},E_{Q'},\alpha_{Q'}\cup \sigma,\theta_{Q'}\cup(n',\theta_n\wedge\varphi))\ :\ \sigma\in \{+,-\}^{\mathcal{N}_Q(n)}\}$$
donde $Q'=Cl_Q^{\mathcal{N}_Q(n)}$, y $\{+,-\}^{\mathcal{N}_Q(n)}$ es el conjunto todas las posibles asignaciones de signo a los elementos de $\mathcal{N}_Q(n)$ (el entorno, en $Q$, del nodo $n$).

Entonces $Q+\{n\wedge \varphi\}$ es un conjunto de refinamiento de $Q$ en $G$ (Fig. \ref{ref4}).
\end{theorem}

\begin{proof}
    La demostración es similar a la realizada en los casos anteriores. Solo hay que tener en cuenta que, cuando se modifica el nodo $n$, no solo queda modificado el $Q$-predicado asociado a él sino también el de todos sus nodos adyacentes.

    Por ello, el procedimiento que se ha seguido para cubrir todas las posibles opciones de asignación de signos para los nodos involucrados es por medio del conjunto de funciones $\{+,-\}^{\mathcal{N}_Q(n)}$ (recordemos que en $\mathcal{N}_Q(n)$ también se tiene en cuenta el centro, $n$).
\end{proof}

Se debe tener en cuenta que los refinamientos anteriores generan estructuras que pueden ser simplificadas. A continuación vamos a definir la operación principal que permite simplificar un GGQ determinado obteniendo otro equivalente con menor número de elementos.

\begin{definition}
    Dado $Q\in GGQ$, diremos que $Q'\subseteq Q$ es \emph{redundante} en $Q$ si $Q\equiv Q-Q'$. Donde $Q-Q'$ es el subgrafo de $Q$ dado por:

    $$(V_Q\smallsetminus V_{Q'}, E_Q\smallsetminus (E_{Q'}\cup\{\gamma(n):\ n\in V_{Q'}\}),\mu_Q)$$
\end{definition}

Veamos un primer resultado que, analizando nodos, nos permite obtener versiones simplificadas de un GGQ por medio de la eliminación de nodos redundantes positivos:

\begin{theorem}
    Sea $Q\in GGQ$, y $n\in V_Q^+$ tal que existe $m\in V_Q$ verificando:
    \begin{itemize}
        \item $\alpha(n)=\alpha(m)$, $\theta_n\equiv\theta_m$.
        \item Para cada $e\in \gamma(n)$, existe $e'\in \gamma(m)$, verificando $\alpha(e)=\alpha(e')$, $\theta_e=\theta_{e'}$ y $\gamma(e)\smallsetminus\{n\}=\gamma(e')\smallsetminus\{m\}$.
    \end{itemize}
    Entonces, $n$ es redundante en $Q$.
\end{theorem}

Esencialmente, la condición que impone el resultado anterior es que $m$ sea un clon de $n$ pero, posiblemente, con más aristas conectadas. Teniendo en mente esta idea intuitiva, la prueba es directa a partir de las condiciones impuestas.

Podemos obtener un resultado similar para aristas por medio del siguiente resultado:

\begin{theorem}
    Sea $Q\in GGQ$, y dos aristas, $e, e'\in E_Q$, tales que $n^+\overset {e}{\longrightarrow} m^+$ y $n^+\overset {e'}{\longrightarrow} m^+$. Si $\theta_e\rightarrow \theta_{e'}$ entonces $e'$ es redundante en $Q$.
\end{theorem}

\begin{figure}[h]
    \begin{center}
        \includegraphics[scale=0.30]{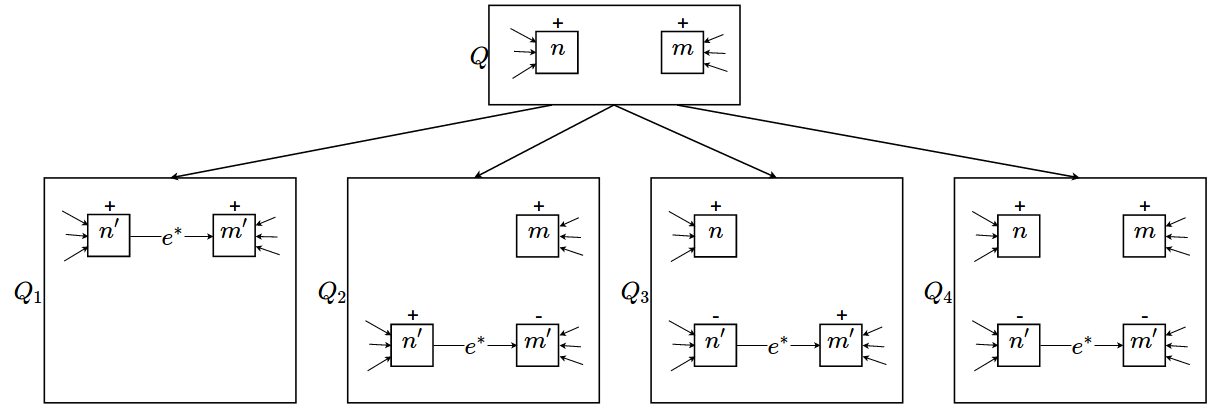}
    \end{center}
    \caption{%
        Refinamiento añadir arista (simplificado).
    }%
    \label{ref2sim}
\end{figure}

\begin{figure}[h]
    \begin{center}
        \includegraphics[scale=0.30]{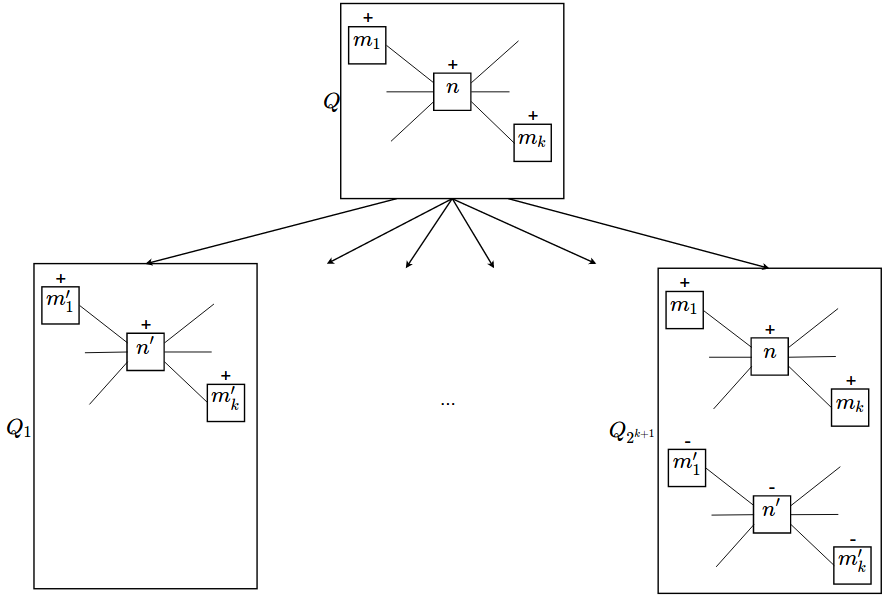}
    \end{center}
    \caption{%
        Refinamiento añadir predicado a nodo (simplificado).
    }%
    \label{ref4sim}
\end{figure}

A partir de los resultados anteriores podemos dar versiones simplificadas de los conjuntos de refinamiento vistos, agrupando nodos positivos y aristas positivas en aquellos casos en los que, tras la clonación inicial, el signo del elemento duplicado se ha mantenido con el original, así como en los casos en los que el signo se ha mantenido y se ha añadido un predicado adicional. En las Figuras \ref{ref2sim} a \ref{ref4sim} se muestran diagramas de los conjuntos de refinamiento $Q+\{n\wedge \varphi\}$, $Q+\{n^+\overset {e \wedge \varphi}{\longrightarrow} m^+\}$ y $Q+\{n\wedge \varphi\}$, respectivamente, aplicando las simplificaciones presentadas.

\begin{figure}[h]
    \begin{center}
        \includegraphics[scale=0.30]{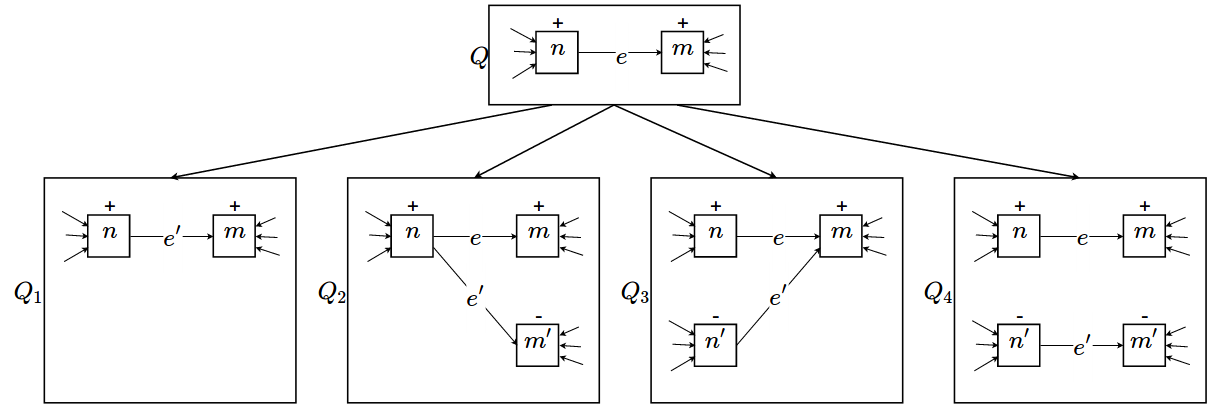}
    \end{center}
    \caption{%
        Refinamiento añadir predicado a arista (simplificado).
    }%
    \label{ref3sim}
\end{figure}

\begin{figure}[h]
    \begin{center}
        \includegraphics[scale=0.35]{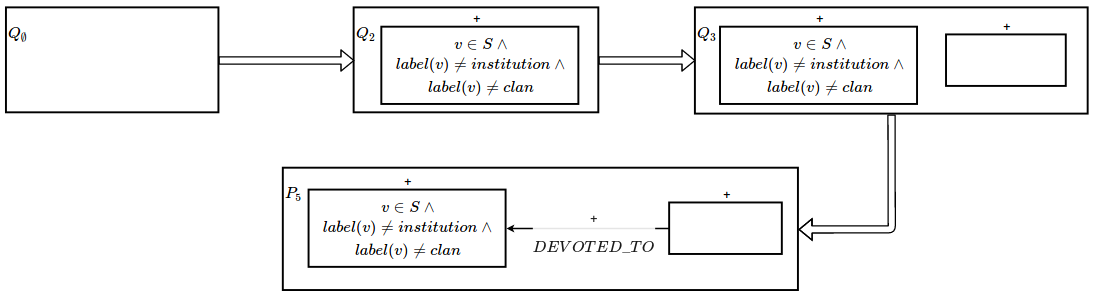}
    \end{center}
    \caption{%
        Sucesión de refinamientos para $P_5$.
    }%
    \label{pqg5refs}
\end{figure}

Por ejemplo, para construir el patrón $P_5$ una posibilidad sería seguir la siguiente secuencia de refinamientos (Fig. \ref{pqg5refs}):
\begin{align*}
Q_1 &= Q_\emptyset+\{n_1\}\\
Q_2 &= Q_1+\{n_1 \land (v \in S \land \tau(v) \neq \mathtt{institution} \land \tau(v) \neq \mathtt{clan} \}\\
Q_3 &= Q_2+\{n_2\}\\
Q_4 &= Q_3+\{n_2 \overset{e_1}{\longrightarrow} n_1\}\\
P_5 &= Q_4+\{n_2 \overset{e_1 \land (\tau(\rho) = \mathtt{DEVOTED\_TO})}{\longrightarrow}n_1\}
\end{align*}

\begin{figure}[h]
    \begin{center}
        \includegraphics[scale=0.25]{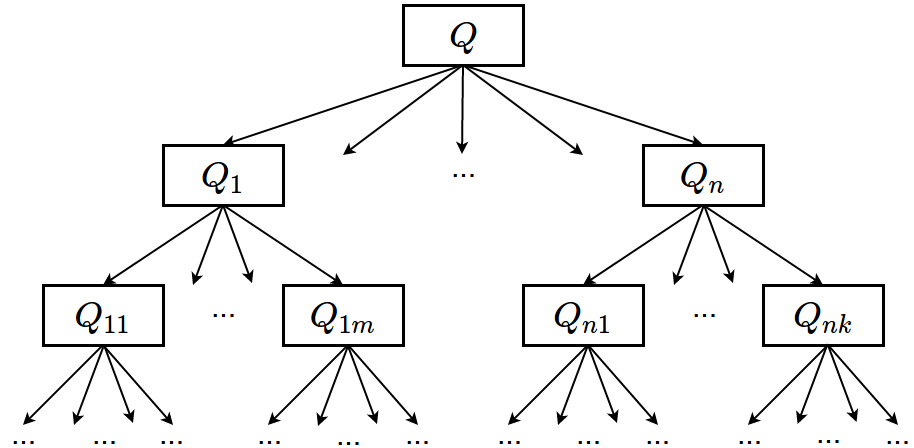}
    \end{center}
    \caption{%
        Árbol de refinamientos.
    }%
    \label{arbolPQG}
\end{figure}

A partir de la estructura de un GGQ no es fácil obtener un GGQ complementario con él. Sin embargo, hay muchos procesos de análisis sobre grafos con propiedades en los que necesitamos trabajar con sucesiones de consultas que verifiquen algunas propiedades de contención y complementariedad como predicados. Los refinamientos vistos en esta sección vienen a cubrir esta carencia y permiten, por ejemplo, construir un árbol de particiones encajadas con los nodos etiquetados de la siguiente forma (Fig. \ref{arbolPQG}):
\begin{itemize}
    \item El nodo raíz está etiquetado con $Q_0$ (un GGQ inicial cualquiera).
    \item Si un nodo del árbol está etiquetado con $Q$, y $R=(Q_1,\dots,Q_n)$ es un conjunto de refinamiento de $Q$, entonces sus nodos hijo se etiquetan con los elementos de $R$.
\end{itemize}

Obsérvese que la construcción del árbol anterior depende por completo de la elección del conjunto de refinamiento que se elija en cada ramificación. 

Los refinamientos que hemos presentado en los resultados anteriores son una opción, pero no es la única posible. Por ejemplo, se pueden considerar refinamientos que, en vez de añadir restricciones a elementos positivos, aligeren las condiciones impuestas por los elementos negativos, consiguiendo nuevos GGQ que refinan al anterior, y usando la adición de predicados por medio de la disyunción en vez de la conjunción.

\section{Conclusiones y Trabajo Futuro}

En este trabajo hemos abordado el objetivo de obtener una herramienta para evaluar subgrafos inmersos en grafos con propiedades de manera que pueda ser utilizada en procedimientos de descubrimiento de información relacional. Para conseguir una herramienta de este tipo era deseable verificar varios requisitos:
\begin{itemize}
	\item Por una parte, resultaba necesario disponer de una gramática que expresase las consultas a evaluar de una forma cercana a las propias estructuras sobre las que iba a trabajar. Gracias a la capacidad expresiva de los grafos generalizados hemos presentado una herramienta de consulta que se puede expresar de forma natural por medio de un grafo con propiedades. 
	\item Además, era necesario dotar al sistema de consulta una base bien fundamentada de propiedades que nos asegurasen que, al ser usadas como predicados lógicos sobre grafos, se comportaban de manera coherente y robusta. Este resultado se ha obtenido presentando las relaciones existentes entre la estructura topológica de la consulta y las relaciones de implicación por medio del refinado. 
	\item Además, era necesario, ya que en también las usaremos para generar métodos automáticos de aprendizaje, que las consultas pudiesen ser modificadas de manera controlada por medio de operadores atómicos que tradujesen el control topológico en un control lógico. En este sentido, se ha  introducido una primera familia de refinamientos que permiten construir a partir de una consulta inicial una colección ordenada de consultas que recorren las diversas opciones de verificación, formando un retículo completo de consultas.
\end{itemize}

Debido a que cualquier estructura de datos relacional puede ser vista como un grafo, y cualquier consulta puede ser vista como la búsqueda de un patrón, la mayoría de lenguajes de consulta en bases de datos pueden ser vistos como herramientas (quizás primitivas) de consulta de patrones en grafos con propiedades. En este trabajo también se han analizado algunas de las herramientas de consulta existentes, así como la viabilidad para ser utilizadas en procedimientos automáticos. Una de las herramientas analizadas, los grafos de selección, permite evaluar registros en bases de datos relacionales a través de patrones acíclicos que pueden ser refinados a partir de operaciones básicas, permitiendo obtener patrones complementarios en cada caso. Para ello, no requiere una proyección exacta del patrón que representa el grafo de selección sobre el subgrafo a evaluar, sino el cumplimiento de una serie de predicados expresados a través de dicho patrón. Debemos recordar que si se exige una proyección a la hora de realizar la verificación de un patrón  se complica la tarea de evaluar la no existencia de determinados elementos. Concretamente, los grafos de selección, evalúan la existencia / no existencia de caminos incidentes al registro bajo evaluación (solo son capaces de evaluar registros individuales), para ello se verifica si se cumple una conjunción de predicados sobre caminos que parten del registro analizado, lo cual puede ser visto como la evaluación de existencia de un árbol enraizado en el nodo que representa el registro bajo evaluación.

Los Generalized Graph Queries que hemos presentado aquí extienden el concepto de grafo de selección permitiendo la evaluación de subgrafos generales, más allá de un único nodo, y el uso de predicados abiertos a través de la definición de un lenguaje sobre los elementos del grafo y patrones cíclicos. Como se convierte en un requisito no usar una proyección para la verificación de un patrón, estos objetivos los hemos conseguido extendiendo la forma de evaluación, que puede ser vista como la evaluación de un árbol enraizado por cada nodo presente en el patrón. A pesar de que por cada nodo de un GGQ se evalúa la existencia de un nodo que cumpla con las condiciones impuestas por su predicado y las aristas en las que participa, al permitir que las aristas se identifiquen con caminos en el grafo (Regular Pattern Matching) se produce la evaluación de un árbol por cada nodo, y no de un simple árbol. Las intersecciones que se producen entre los diversos árboles y las restricciones impuestas en los nodos permiten la evaluación de patrones cíclicos en los GGQ, algo que no se había conseguido en otras propuestas anteriores.

Como hemos comentado, al igual que los grafos de selección, los GGQ se pueden modificar y construir  a partir de refinamientos, pero a diferencia del caso simple de los grafos de selección, normalmente los refinamientos no son binarios, ya que su aplicación puede modificar más de un predicado en el patrón, dando lugar a conjuntos de tamaño $2^k$ (siendo $k$ el número de predicados modificados). A través de la definición de determinadas operaciones de simplificación y equivalencia, los refinamientos mostrados pueden ser simplificados dando lugar a herramientas sencillas que permiten construir consultas complejas en grafos. 

En general, los refinamientos dan lugar a particiones encajadas de las estructuras que evalúan, lo que los convierte en herramientas ideales para procedimientos de caja blanca. Tras haber llevado a cabo una primera implementación como prueba de concepto (pero totalmente funcional), se ha demostrado experimentalmente que los GGQ son viables bajo condiciones suaves y que cumplen con los objetivos planteados de extensión de las herramientas existentes. 

Un uso explícito de estas capacidades ya ha sido llevado a cabo en procedimientos de descubrimiento de información, en concreto en el algoritmo GGQ-ID3, que hace uso de los Generalized Graph Queries como herramientas de test para la construcción de un árbol de decisión siguiendo los fundamentos del famoso algoritmo ID3. La relación que guardan los GGQ con GGQ-ID3 es equivalente a la relación que guardan los grafos de selección con el algoritmo MRDTL \cite{Leiva02mrdtl:a}. En los resultados de los experimentos llevados a cabo, se muestra que GGQ-ID3 es capaz de extraer patrones interesantes que pueden ser utilizados en tareas de aprendizaje complejas.

Por otro lado, se pueden crear familias de refinamientos más complejos (por ejemplo, combinar el refinamiento \emph{añadir arista} con \emph{añadir propiedad a una arista} en un solo paso) para de esta manera reducir el número de pasos para obtener GGQ complejos y ampliar la potencia con respecto a los pasos atómicos que son menos informativos. Si se lleva a cabo esta opción de manera adecuada (unificando los refinamientos en función de la frecuencia de aparición de estructuras en un grafo, por ejemplo) se puede conseguir que los algoritmos de descubrimiento que hacen uso de GGQ se acerquen de manera más rápida a una buena solución. En este caso se consigue una mejora en la eficiencia sacrificando la posibilidad de cubrir un espacio de consultas más amplio. En este sentido, en este trabajo se ha ofrecido un conjunto minimal pero robusto de refinamientos, pero debe tenerse en cuenta que no se ofrecen con la intención de que sea óptimo para ciertas tareas de aprendizaje.

En consecuencia, los modelos de consulta en grafos basados en GGQ permiten obtener herramientas potentes y sencillas, de complejidad controlada, idóneas para su construcción automática y para ser utilizadas en tareas de caja blanca sobre información multi-relacional, debido en parte a las buenas propiedades relacionadas con complementariedad y contención de consultas. 

Con respecto a los trabajos futuros que derivan del desarrollo aquí presentado, cabe mencionar que, gracias a que los GGQ están construidos utilizando la estructura de grafo generalizado, y que dicha estructura permite la definición de hipergrafos de manera natural, los GGQ pueden evaluar hipergrafos con propiedades teniendo en cuenta pequeñas modificaciones sobre las definiciones presentadas, por lo que la extensión de los Generalized Graph Queries hacia Generalized Hypergraph Queries es un paso natural que merece la pena ser considerado. Además, el desarrollo de diferentes conjuntos de refinamiento en función del tipo de grafo a consultar o incluso la generación automática de dichos conjuntos a partir de estadísticas extraídas del grafo a analizar puede derivar en optimizaciones importantes en procesos de construcción automática y efectiva de GGQ. Por útlimo, cabe destacar que los GGQ ya están siendo utilizados por procedimientos de descubrimiento/aprendizaje como el algoritmo GGQ-ID3 de construcción de árboles multi-relacionales nombrado anteriormente, pero gracias a sus buenas propiedades, son grandes candidatos para ser utilizados por otros algoritmos de este tipo. 

\bibliographystyle{plain}	% or "siam", or "alpha", etc.
\bibliography{biblio}